\documentclass[twocolumn]{aastex63}
\pdfoutput=1 
\usepackage{amsmath,amstext}
\usepackage[T1]{fontenc}
\usepackage{apjfonts} 
\usepackage[figure,figure*]{hypcap}
\usepackage{longtable}
\usepackage{threeparttable}

\usepackage{amssymb}
\usepackage{amsmath}
\usepackage{fontawesome}
\usepackage{gensymb}
\usepackage{mathrsfs}
\usepackage{upgreek}
\usepackage{fontenc}
\usepackage{hyperref}
\usepackage{float}
\usepackage{tabularx}

\newcommand{\ergs}{\text{erg s}\ensuremath{^{-1}}}
\newcommand{\ergscm}{\text{erg s}\ensuremath{^{-1}}\text{ cm}\ensuremath{^{-2}}}

\newcommand{\kms}{\text{km s}\ensuremath{^{-1}}}
\newcommand{\voff}{\ensuremath{v_{\rm off}}}

\renewcommand{\d}{\ensuremath{\rm d}}

\graphicspath{{./}{figures/}}

\begin{document}
\title{The CLASS Quasar Catalog: Coronal Line Activity in Type~1 SDSS Quasars}
\author[0000-0003-3152-4328]{Sara Doan}
 \author[0000-0003-2277-2354]{Shobita Satyapal}
 \affiliation{George Mason University, Department of Physics and Astronomy, MS3F3, 4400 University Drive, Fairfax, VA 22030, USA}
 \author[0000-0003-4701-8497]{Michael Reefe}
\altaffiliation{National Science Foundation, Graduate Research Fellow}
\affiliation{MIT Kavli Institute for Astrophysics and Space Research, Massachusetts Institute of Technology, Cambridge, MA 02139, USA}
 \author[0000-0003-3432-2094]{Remington O. Sexton}
\affiliation{U.S. Naval Observatory, 3450 Massachusetts Avenue NW, Washington, DC 20392-5420, USA}
\author[0000-0003-3937-562X]{William Matzko}
\affiliation{George Mason University, Department of Physics and Astronomy, MS3F3, 4400 University Drive, Fairfax, VA 22030, USA}
\author[0000-0002-0913-3729]{Jeffrey D. McKaig}
\affiliation{George Mason University, Department of Physics and Astronomy, MS3F3, 4400 University Drive, Fairfax, VA 22030, USA}
\author[0000-0002-4902-8077]{Nathan J. Secrest}
\affiliation{U.S. Naval Observatory, 3450 Massachusetts Avenue NW, Washington, DC 20392-5420, USA}
 \author[0000-0003-1051-6564]{Jenna M. Cann}
\altaffiliation{NASA Postdoctoral Program}
\affiliation{Center for Space Sciences and Technology, University of Maryland, Baltimore County, 1000 Hilltop Circle, Baltimore, MD 21250, USA}
\affiliation{X-ray Astrophysics Laboratory, NASA Goddard Space Flight Center, Greenbelt, MD 20771, USA}
\author{Ari Laor}
\affiliation{Physics Department, Technion – Israel Institute of Technology, Haifa 32000, Israel}

\author[0000-0003-4693-6157]{Gabriela Canalizo}
\affiliation{Department of Physics and Astronomy, University of California, Riverside, 900 University Avenue, Riverside, CA 92521, USA}

\begin{abstract}

We conduct the first systematic survey of a total of eleven optical coronal lines in the spectra of a large sample of low redshift (z < 0.8) Type~1 quasars observed by the Sloan Digital Sky Survey (SDSS). We find that strong coronal line emission is rare in SDSS even in Type~1 quasars; only 885 out of 19,508 (4.5\%) galaxies show at least one coronal line, with higher ionization potential lines ($>100$eV) being even rarer. The [\ion{Ne}{5}] $\lambda$3426 line, which constitutes the majority of detections, is strongly correlated with the bolometric luminosity. These findings suggest that the optical coronal lines are significantly suppressed in the majority of local AGNs, possibly as a result of the presence of dust in the emitting regions. We find that the incidence of ionized outflows is significantly higher in coronal line emitters compared with non-coronal line emitters, possibly suggesting that dust destruction in outflows enhances coronal line emission in AGNs. Many coronal lines show line profiles that are broader than those of narrow lines, and are blue-shifted relative the lower ionization potential lines, suggesting outflows in the highly ionized gas. Given the limited number of detections, we do not find any statistically significant trends of detection statistics, or line ratios with black hole mass, Eddington ratio, or AGN bolometric luminosity. The catalog is publicly available and can provide a useful database of the coronal line properties of low redshift quasars that can be compared to the growing number of high-z AGNs discovered by JWST.
\par

\end{abstract}

\section{Introduction}

Observations of high ionization fine-structure lines, often referred to as ``coronal lines'' (CLs) because of their first discovery in the solar corona, are a reliable tool in finding AGNs and characterizing their properties.  This is because even hot massive stars do not produce photons with energies sufficient to produce the ions, and CL emission from Type~II supernovae is extremely weak and short lived \cite[e.g.,][]{1990AJ....100.1588B,2009ApJ...695.1334S}. CLs therefore have the potential to reveal elusive AGNs that are missed by other widely used diagnostics that suffer from obscuration or contamination from star formation in the host galaxy \cite[see][]{2021ApJ...906...35S}. Because of their high ionization potentials and high critical densities ($\sim 10^{7}$--$10^{10}$ cm$^{-3}$), CLs are powerful probes of the highly ionized and dense gas in galaxy centers. The CL emission can extend several hundred parsecs from the nucleus \citep[e.g.,][]{2010MNRAS.405.1315M, 2018ApJ...858...48M, 2021ApJ...920...62N, 2023ApJ...945..127N}, with some of the emission originating between the broad line region and the narrow line region, as demonstrated by recent VLT/GRAVITY observations \citep{2021A&A...648A.117G}.

Not only do CLs have the potential to identify elusive AGNs missed by optical narrow line ratios and mid-infrared color selection, which suffer from contamination from star formation in the host galaxy \cite[e.g.,][]{2021ApJ...906...35S}, they may also be a powerful tool for constraining the black hole accretion properties. Due to their ionization energies, coronal lines offer a method by which to indirectly probe the spectral energy distribution (SED) of the ionizing radiation field from the Lyman limit up to several hundred electron volts, a region of the electromagnetic spectrum that is observationally inaccessible because of Galactic and intrinsic absorption \cite[e.g.,][]{1996A&A...315L.109M,2000ApJ...536..710A}. As a result, they can potentially be used to constrain accretion disk models, providing clues into the physics of accretion and AGN fueling. Because the shape of the ionizing radiation field hardens with decreasing black hole mass, CL flux ratios may even possibly provide constraints on the mass of the central black hole \citep{2018ApJ...861..142C, 2022MNRAS.510.1010P}, thereby offering a powerful tool for identifying accreting intermediate mass black holes \cite[IMBHS; e.g.,][]{2018ApJ...861..142C, 2021ApJ...912L...2C, 2021ApJ...922..155M, 2022ApJ...936..140R,2023ApJ...946L..38R, 2023arXiv230502189H}. The CL spectrum can therefore be a powerful tool for constraining the AGN properties in galaxies where other tools are ineffective. Indeed, black hole masses and Eddington ratios are currently only available for bright Type~1 AGNs, which comprise only a small fraction of the total AGN population. Even amongst the population of AGNs in current surveys based on the strong optical narrow lines, Type~2 AGNs (i.e. the AGNs without visible broad lines) are about 4 times more numerous than Type~1 AGNs \citep{1995ApJ...454...95M}, and Type~1 and 2 AGNs in current optical surveys themselves capture only a small fraction of the total active galaxy population. The black hole masses and accretion properties of the vast majority of accreting black holes in galaxy centers are therefore currently unknown.

Although CLs have been known to exist in the optical spectra of nearby AGNs for almost sixty years \cite[e.g.,][]{1968ApJ...151..807O,1970ApJ...161..811N,1978ApJ...221..501G,1984MNRAS.211P..33P,1988AJ.....95...45A,2002MNRAS.329..309P}, their full diagnostic potential has not been tapped. Recent systematic surveys of the optical CLs in large samples of galaxies are only just emerging, and are revealing that optical CL emission is extremely rare in the general galaxy population \citep{10.1111/j.1365-2966.2009.14961.x, 2022ApJ...936..140R, 2023ApJS..265...21R,2021ApJ...920...62N,2023ApJ...945..127N} given current survey sensitivities, and are preferentially found in galaxies with the least dust extinction \citep{2023ApJ...945..127N}, suggesting that dust plays a role in suppressing the emission. Remarkably, recent optical surveys are revealing CLs with the highest ionization potentials in a population of dwarf galaxies with no other signs of accretion activity \citep{2021ApJ...922..155M,2022ApJ...936..140R,2023ApJS..265...21R,2023ApJ...946L..38R,2023arXiv230502189H}, hinting at the possibility that they are revealing accreting IMBHs in a completely uncharted mass regime. However, these studies are based on surveys of the general galaxy population, the majority of which are not robustly identified AGNs based on standard diagnostics, and fundamental black hole and accretion properties are unavailable. A systematic large-scale investigation of the diagnostic potential of coronal lines in constraining AGN properties requires a systematic large-scale study of confirmed AGNs with broad lines in which black hole masses, AGN bolometric luminosities, and Eddington ratios are available.

In this work, we investigate for the first time the prevalence and properties of optical CL emission using the SDSS spectra of a large sample of Type~1 quasars. We choose Type 1 objects so we can explore CL properties with black hole mass and accretion properties. Following the methodology adopted in previous papers aimed at the general galaxy population \citep{2022ApJ...936..140R,2023ApJS..265...21R}, we introduce in this work the Coronal Line Activity Spectroscopic Survey (CLASS) SDSS quasar catalog (CLASS-Q). We quantify for the first time the detection statistics for 11 optical CLs detected in the SDSS spectra of a large sample of Type~1 quasars and explore their dependence on black hole mass and AGN properties. The CLASS-Q catalog, with  all CL fluxes and profiles, will be made publicly available through Vizier. We also search for and characterize the properties of outflows in the [\ion{O}{3}]~$\lambda$5007 in the full sample to explore trends in CL incidence and the presence and properties of outflows in the large-scale ionized gas traced by the [\ion{O}{3}]~$\lambda$5007 line. The results from this paper may be useful in predicting CL fluxes in high redshift galaxies observed by JWST. Throughout this paper, we assume a flat $\Lambda$CDM cosmology with $H_0=70$~km~s$^{-1}$~Mpc$^{-1}$, $\Omega_m = 0.3$, and $\Omega_\Lambda = 0.7$.

\section{Sample Selection and Methodology}
\begin{figure}
    \centering 
    \includegraphics[width=\columnwidth]{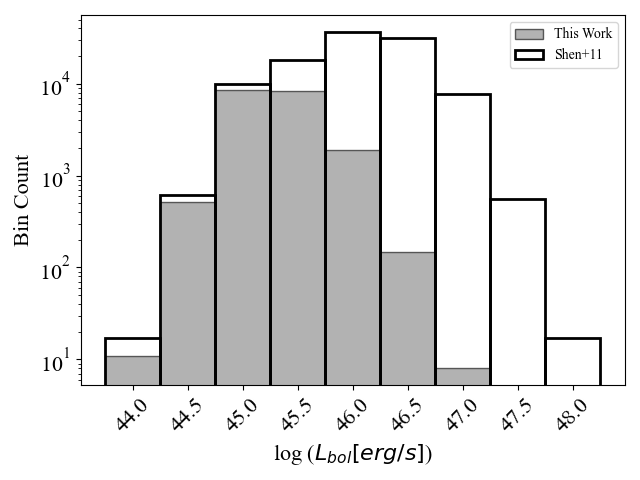}
    \includegraphics[width=\columnwidth]{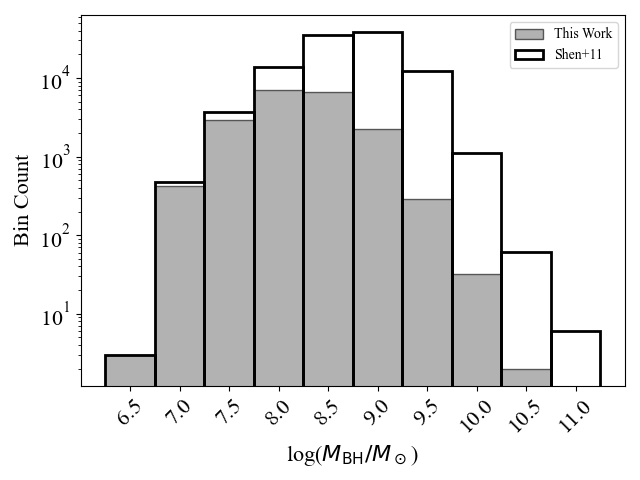}
    \includegraphics[width=\columnwidth]{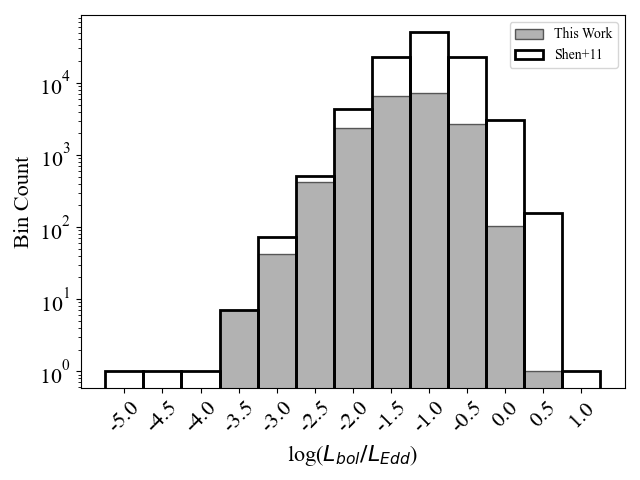}
    \caption{Bolometric luminosity, black hole mass, and Eddington ratio distributions of our sample compared with the parent sample of quasars from \citet{shen2011}. Quantities displayed are taken directly from the catalog values provided by \citet{shen2011}.}
    \label{fig:sample_histo1}
\end{figure}

 Our sample of Type~1 AGNs is drawn from the spectroscopic catalog of quasars from SDSS~DR7 compiled by \citet{shen2011}, which contains contains 105,783 bona fide quasars brighter than $M_{i} = -22.0$, and has at least one broad emission line with full width at half-maximum (FWHM) larger than 1000~km~s$^{-1}$.  Each quasar from \cite{shen2011} was matched to the SDSS~DR16 quasar catalog in order to use the most accurate available redshift. In order to determine the relative strength of the CL emission with the strong optical narrow lines typically observed in AGNs, we impose a redshift cut of $z < 0.8$ to ensure that the [\ion{O}{3}]\,$\lambda$5007 line is within the redshift range of the SDSS spectrum. This also allows us to search for ionized outflows in the [\ion{O}{3}]\,$\lambda$5007 line, a widely used tracer to search for and characterize large scale outflows in galaxies \citep[e.g.][]{2014MNRAS.441.3306H,2016ApJ...828...97B,2020MNRAS.492.4680W,2022MNRAS.514.4828M, 2024A&A...687A.111M}. Requiring this cut reduces the sample size to 19,508. In Figure~\ref{fig:sample_histo1}, we show the bolometric luminosity, black hole mass, and Eddington ratio distributions of the resulting sample compared with the parent sample. Our imposed redshift cut results in a sample that spans a wide range in parameter space, but excludes the most luminous quasars in the parent sample.  

\subsection{Pre-selection filtering}
We follow the methodology adopted in a previous series of papers to search for CL detections in the sample \citep{2022ApJ...936..140R, 2023ApJS..265...21R}. We begin with a pre-selection strategy to identify candidate CL emitters in the starting sample of 19,508 objects in order to reduce computation time and eliminate spurious detections. This methodology is designed to identify robust detections, and will necessarily exclude marginal detections. We refer the reader to the previous papers for full details.  In brief, for each of the CLs listed in Table \ref{tab:line_props}, we perform a linear fit to the continuum using two adjacent wavelength windows $\pm$30 \AA\ on either side of the the line of interest, adjusting the location as necessary to ensure that they fall on a flat and featureless part of the continuum. We subtract this continuum from the average flux in a 20 \AA\ region centered on the CL. We then compare the resulting flux to the root-mean-square (RMS) deviation of the flux in the adjacent reference windows, and require that the average flux centered on each CL is at least $4\sigma$ above the noise. As in previous papers, this filtering is achieved with a custom-made publicly available Python package called BIFR\"OST\footnote{\href{https://github.com/Michael-Reefe/bifrost}{https://github.com/Michael-Reefe/bifrost}}. As in previous work, in order to eliminate spurious detections, we require that at least 3 continuous pixels are $3\sigma$ above the continuum for each line, and that they are not within $\pm$20 \AA\ (or $\sim \pm 1000$ \kms) of the four most prominent sky lines at 5578.5, 5894.6, 6301.7, and 7246.0 \AA\ respectively.  We note that we eliminate all CLs with wavelengths in proximity to the regions in the continuum dominated by the complex of blended \ion{Fe}{2} multiplets (for a review, see \citet{2022AN....34310112G}, and references therein)  which are prevalent in the spectra of Type~1 quasars. These features, and the variations in their presence and strength in the spectra of quasars, results in significant challenges in fitting the continuum and can thus significantly impact the spectral line fits and resulting fit parameters in a non-uniform way \cite[e.g.,][]{2023A&A...679A..34P}. Imposing this restriction eliminated 9 optical CLs in the SDSS wavelength regime. The resulting set of 11 CLs free of contamination that are  included in this work are listed in Table~\ref{tab:line_props}. After all pre-selection filtering was completed, the resulting number of CL candidate detections was 1092, in 953 unique targets.

\subsection{Fitting}
Spectral fitting was performed on all candidate detections and is done, as in previous work, with the open-source Python code Bayesian AGN Decomposition Analysis for SDSS Spectra (\textsc{badass})\footnote{\href{https://github.com/remingtonsexton/BADASS3}{https://github.com/remingtonsexton/BADASS3}} (customized version based on v9.2.2). A full description of the code can be found in \citet[][]{sexton_2020}. Spectra are fit with BADASS first using a likelihood maximization routine from SciPy, followed by a Markov Chain Monte Carlo (MCMC) fit using the affine-invariant Emcee sampler \citep{2013PASP..125..306F} which yields robust uncertainties and covariances.  BADASS models the AGN power-law continuum, \ion{Fe}{2} emission, and spectral lines, simultaneously with the stellar line-of-sight velocity distribution (LOSVD) using the penalized pixel-fitting \cite[pPXF;][]{2017MNRAS.466..798C} method.  Each CL is modeled as a simple Gaussian profile with a free amplitude, velocity offset, and width, with outflow and stellar continuum fitting conducted as outlined in \cite{sexton_2020}. A detailed description of our fitting procedure can be found in previous work \citep{2022ApJ...936..140R, 2023ApJS..265...21R}. 

After fitting each coronal line, we require an equivalent width cut of 1.0 \AA\ and a flux cut of $10^{-17} \ergscm$, to ensure a robust detection given the sensitivity limit of SDSS. We then perform a final round of visual inspection to confirm each detection.
Our final sample contains 885 quasars from the starting sample of 19,508 objects, and a total of 974 CL detections. We present a selection of individual spectra that show some of the most commonly detected coronal lines in Figure \ref{fig:spectra}, with the spectral fit overlaid. Note that we carefully examined all detections of the [\ion{Fe}{10}] $\lambda$6374 line and ensured that there is no contamination of this line from the \ion{Si}{2} $\lambda$6371 line, as has been reported in a dwarf galaxy in recent work \citep{2023RNAAS...7...99H}. None of our targets reveal the stronger \ion{Si}{2} $\lambda$6347 line in their SDSS spectra. Given also that our sample are exclusively robustly identified AGNs, it is likely that none of the 7 [\ion{Fe}{10}] $\lambda$6374 lines reported in this work are misidentified \ion{Si}{2} $\lambda$6371 lines.

\begin{figure*}
    \centering
    \includegraphics[width=\columnwidth]{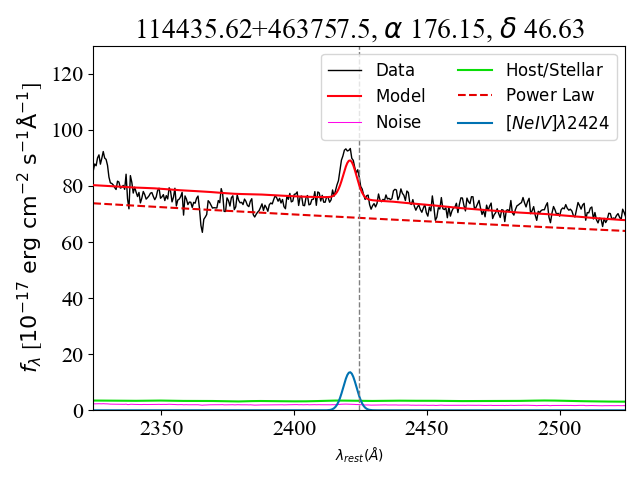}
    \includegraphics[width=\columnwidth]{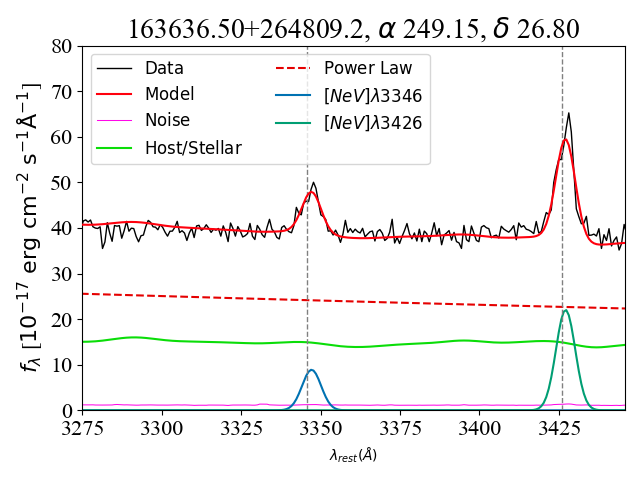}
    \includegraphics[width=\columnwidth]{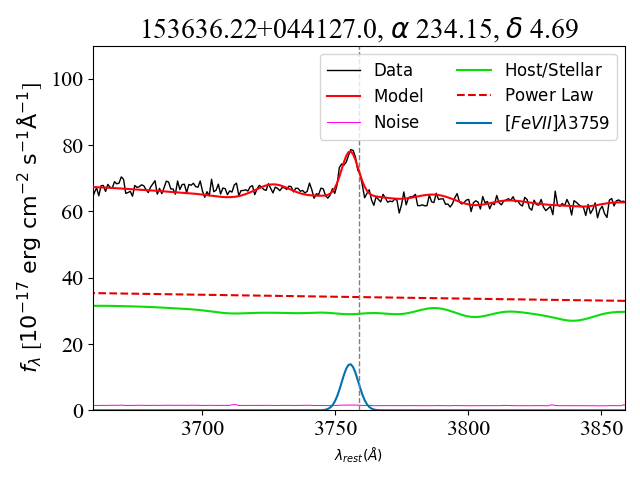}
    \includegraphics[width=\columnwidth]{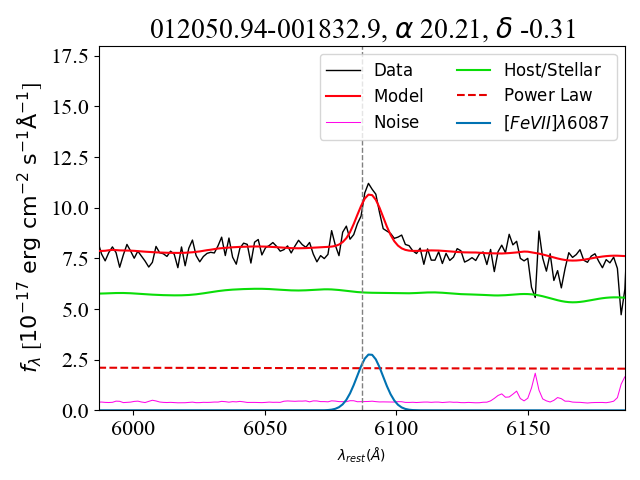}
    \includegraphics[width=\columnwidth]{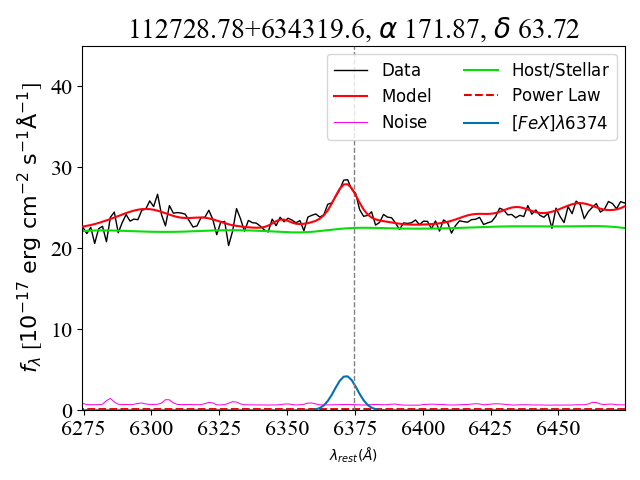}
    \includegraphics[width=\columnwidth]{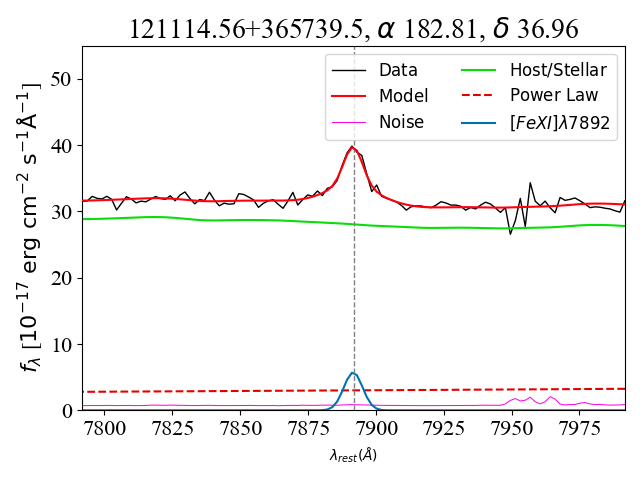}
    
    \caption{Examples of 6 individual spectra that exhibit emission of at least one coronal line in the optical.  The SDSS data, corrected for redshift and galactic extinction, are plotted in black, while the model and each of its components (AGN power law, host galaxy, and emission lines) are overlaid on the spectra.  Each coronal line is labeled, and the object's SDSS Spec Object ID and coordinates are labeled in each plot's title.}
    \label{fig:spectra}
\end{figure*}

In Table \ref{tab:detect_stats}, we list detection statistics in the sample for each CL. We determine the detection fraction following the same methodology applied by \citet{2023ApJS..265...21R}. In brief, in calculating a detection fraction, we account for variations in signal-to-noise of spectra by excluding observations in which the sensitivity is not sufficient to enable a 3$\sigma$ detection of the given CL with a luminosity equal to the average luminosity (column 8) or the maximum luminosity (column 9) of the CL amongst our detections. This allows us to account for the non-uniform sensitivity of the observations to each of the CLs, allowing their detection percentages to be directly comparable to each other, given the sensitivity of SDSS. In addition, we exclude in our calculation of detection fraction observations in which the given coronal line is redshifted out of the SDSS spectral range. We adopt this uniform procedure in calculating detection fraction so we can readily compare the statistics obtained in the CLASS-Q catalog presented here, with those presented in the general CLASS galaxy catalog \citep{2022ApJ...936..140R,2023ApJS..265...21R}. Note that detection fractions also depend on the strength of the continuum; the reported detection fractions are therefore intended to report a fraction based on a simple methodology, intended to illustrate the rarity of detections at the sensitivity limit of SDSS, and to allow direct comparison to the detection fractions reported in the previously published CLASS catalog. The catalog lists all objects in the sample, with luminosities of all CLs for detections and threshold luminosities. This will allow detection fractions to be obtained through alternative methodologies.


\begin{deluxetable}{ccccc}
\tabletypesize{\footnotesize}
\tablecaption{Optical Coronal Lines. $\lambda_{\rm rest}$ is the rest wavelength, $\rho_{\rm crit}$ is the critical density, IP is the ionization potential.}
\tablehead{\colhead{Line} & \colhead{$\lambda_{\rm rest}$}\textsuperscript{1} & \colhead{$\rho_{\rm crit}$} & \colhead{IP}\textsuperscript{2} & \colhead{Transition}\\
\colhead{} & \colhead{(\AA)} & \colhead{(cm$^{-3}$)} & \colhead{(eV)} & \colhead{}}
\decimals
\startdata
\lbrack \ion{Fe}{11}\rbrack & 7891.800 & $6.39 \times 10^8$ & 262.10 & $^3$P$_2 - ^3$P$_1$ \\
\lbrack \ion{S}{12}\rbrack & 7611.000 & $7.09 \times 10^9$ & 504.78 & $^2$P$^0_{1/2} - ^2$P$^0_{3/2}$ \\
\lbrack \ion{Fe}{10}\rbrack & 6374.510 & $4.45 \times 10^8$ & 235.04 & $^2$P$^0_{3/2} - ^2$P$^0_{1/2}$ \\
\lbrack \ion{Fe}{7}\rbrack & 6087.000 & $4.46 \times 10^7$ & 99.00 & $^3$F$_3 - ^1$D$_2$ \\
\lbrack \ion{Fe}{7}\rbrack & 5720.700 & $3.72 \times 10^7$ & 99.00 & $^3$F$_2 - ^1$D$_2$ \\
\lbrack \ion{Fe}{5}\rbrack & 3891.280 & $1.61 \times 10^8$ & 54.80 & $^5$D$_4 - ^3$F2$_4$ \\
\lbrack \ion{Fe}{5}\rbrack & 3839.270 & $1.00 \times 10^8$ & 54.80 & $^5$D$_3 - ^3$F2$_3$ \\
\lbrack \ion{Fe}{7}\rbrack & 3758.920 & $4.02 \times 10^7$ & 99.00 & $^3$F$_4 - ^1$G$_4$ \\
\lbrack \ion{Ne}{5}\rbrack & 3425.881 & $1.90 \times 10^7$ & 97.11 & $^3$P$_2 - ^1$D$_2$ \\
\lbrack \ion{Ne}{5}\rbrack & 3345.821 & $1.14 \times 10^7$ & 97.11 & $^3$P$_1 - ^1$D$_2$ \\
\lbrack \ion{Ne}{4}\rbrack & 2424.403 & $4.24 \times 10^4$ & 63.42 & $^4$S$^0_{3/2} - ^2$D$^0_{5/2}$ \\
\enddata
\begin{tablenotes}
	\item[0] \textsuperscript{1}Wavelengths taken from: \url{https://physics.nist.gov/PhysRefData/ASD/lines_form.html}.
	\item[1] \textsuperscript{2}Ionization potential taken from: \url{https://physics.nist.gov/PhysRefData/ASD/ionEnergy.html}.
 \medskip
 
 For the NIST database, see Kramida, A., Ralchenko, Yu., Reader, J. and NIST ASD Team (2022). NIST Atomic Spectra Database (version 5.10), [Online]. Available: https://physics.nist.gov/asd [Thu Jun 01 2023]. National Institute of Standards and Technology, Gaithersburg, MD. DOI: https://doi.org/10.18434/T4W30F
 
\end{tablenotes}
\label{tab:line_props}
\end{deluxetable}


\startlongtable
\begin{deluxetable*}{ccccccccccccc}
\tabletypesize{\footnotesize}
\tablecaption{Detection Properties: Detection rates and line fit statistics for the coronal lines in Table \ref{tab:line_props}. $L_{Thresh}$ is the average $1\sigma$ luminosity detection threshold across all spectra, $L$ is the luminosity, FWHM is the full-width at half-maximum of the line profile, $v_{\rm off}$ is the velocity offset relative to the stellar velocity, EW is the equivalent width, and $N$ is the number of detections.The values shown are (mean) $\pm$ (standard deviation).}
\tablehead{\colhead{Line} & \colhead{$\lambda_{\rm rest}$}\textsuperscript{1} & \colhead{$L_{Thresh}$} & \colhead{$L$} & \colhead{FWHM} & \colhead{$v_{\rm off}$} & \colhead{EW} & \colhead{Detections (CL avg)}\textsuperscript{2} & \colhead{Detections (CL max)}\textsuperscript{2} & \colhead{$N$}\\
\colhead{} & \colhead{(\AA)} & \colhead{$\log L/$cgs}  & \colhead{$\log L/$cgs}  & \colhead{(km s$^{-1}$)} & \colhead{(km s$^{-1}$)} & \colhead{(\AA)} & \colhead{(\%)} & \colhead{(\%)} & \colhead{}}
\decimals
\startdata
\lbrack \ion{Fe}{11}\rbrack & 7891.800 & 40.07 & $40.68 \pm 0.00$ & $282 \pm 0$ & $-6 \pm 0$ & $1.60 \pm 0.00$ & ${0.22}^{+0.50}_{-0.18}$ & ${0.22}^{+0.50}_{-0.18}$ & 1 \\
\lbrack \ion{S}{12}\rbrack & 7611.000 & 40.21 & --- & --- & --- & --- & --- & --- & 0 \\
\lbrack \ion{Fe}{10}\rbrack & 6374.510 & 40.43 & $40.73 \pm 0.21$ & $646 \pm 247$ & $-25 \pm 86$ & $2.13 \pm 0.62$ & ${0.180}^{+0.096}_{-0.066}$ & ${0.113}^{+0.061}_{-0.042}$ & 7 \\
\lbrack \ion{Fe}{7}\rbrack & 6087.000 & 40.49 & $40.93 \pm 0.29$ & $617 \pm 208$ & $-50 \pm 117$ & $2.23 \pm 1.04$ & ${1.14}^{+0.15}_{-0.13}$ & ${0.772}^{+0.102}_{-0.091}$ & 71 \\
\lbrack \ion{Fe}{7}\rbrack & 5720.700 & 40.61 & $40.76 \pm 0.19$ & $547 \pm 194$ & $-31 \pm 114$ & $1.94 \pm 0.66$ & ${0.225}^{+0.077}_{-0.059}$ & ${0.164}^{+0.056}_{-0.043}$ & 14 \\
\lbrack \ion{Fe}{5}\rbrack & 3891.280 & 40.69 & $40.88 \pm 0.00$ & $1038 \pm 0$ & $90 \pm 0$ & $2.12 \pm 0.00$ & ${0.017}^{+0.039}_{-0.014}$ & ${0.017}^{+0.039}_{-0.014}$ & 1 \\
\lbrack \ion{Fe}{5}\rbrack & 3839.270 & 40.64 & --- & --- & --- & --- & --- & --- & 0 \\
\lbrack \ion{Fe}{7}\rbrack & 3758.920 & 40.81 & $41.95 \pm 0.42$ & $840 \pm 341$ & $37 \pm 243$ & $4.12 \pm 4.27$ & ${0.0207}^{+0.0163}_{-0.0099}$ & ${0.0205}^{+0.0162}_{-0.0098}$ & 4 \\
\lbrack \ion{Ne}{5}\rbrack & 3425.881 & 40.62 & $41.88 \pm 0.37$ & $787 \pm 274$ & $-107 \pm 115$ & $4.64 \pm 4.17$ & ${4.38}^{+0.15}_{-0.15}$ & ${4.36}^{+0.15}_{-0.15}$ & 847 \\
\lbrack \ion{Ne}{5}\rbrack & 3345.821 & 40.64 & $41.67 \pm 0.31$ & $665 \pm 196$ & $-77 \pm 103$ & $4.38 \pm 3.78$ & ${0.100}^{+0.028}_{-0.023}$ & ${0.098}^{+0.028}_{-0.022}$ & 19 \\
\lbrack \ion{Ne}{4}\rbrack & 2424.403 & 41.09 & $42.47 \pm 0.22$ & $893 \pm 238$ & $-348 \pm 101$ & $1.45 \pm 0.43$ & ${0.122}^{+0.052}_{-0.038}$ & ${0.122}^{+0.052}_{-0.038}$ & 10 \\
\enddata
\begin{tablenotes}
	\item[0] \textsuperscript{1}Wavelengths taken from: \url{https://physics.nist.gov/PhysRefData/ASD/lines_form.html}.
	\item[1] \textsuperscript{2}The detection percentage columns are calculated as the number N over the total number of spectra that \
          could have detected the line. The denominator of the left detection column is determined by the number \
          of galaxies whose luminosity threshold exceeds 3 times the average CL luminosity and the right is determined \
          with respect to 3 times the maximum CL luminosity. In both cases, the errors are 68\% confidence intervals \
          calculated with binomial statistics.
\end{tablenotes}
\label{tab:detect_stats}
\end{deluxetable*}

\subsection{[OIII]~$\lambda$5007 Outflow Fitting}

 We perform spectral fitting of the [OIII]~$\lambda$5007 emission line to search for outflows in the sample of 19,508 in order to explore possible trends in CL detections and the presence of large-scale outflows in the sample. Because this line is one of the strongest features in the optical spectrum of all emission line galaxies, and is produced in the lower density narrow line region, it is commonly used in the literature to search for and characterize outflows \citep[e.g.][]{2014MNRAS.441.3306H, 2016ApJ...828...97B, 2016MNRAS.459.3144Z, 2020MNRAS.492.4680W,2022MNRAS.514.4828M}. In this work, we do not attempt to fit outflow components in the CLs due to their typically lower signal-to-noise spectral fits. To test for the presence of an outflow in [O III] $\lambda$5007, we fit both a 1-component (narrow) and a 2-component (narrow + broad) model to the [O III] doublet and examine the residuals of the resultant fits to determine if the outflowing component is justified. Following \citet{2022MNRAS.514.4828M}, we require the core component and outflow component FWHM and velocity offset parameters to be different at a $0.5\sigma$ level, and the F-test p-value be greater than 0.80. This intentionally low confidence value is used so we do not exclude weak outflows. The final fits obtained with \textsc{emcee} are then analyzed to determine the quality of the outflow fit. In order for the outflow component to be considered robust, we require the final outflow amplitude to be at least 1$\sigma$ above the noise floor of the spectrum, and the FWHM of the outflowing component must be greater than the FWHM of the narrow component within their 1$\sigma$ uncertainties. Figure ~\ref{fig:spectra} shows example fits to the [OIII]~$\lambda$5007 line in which outflow components were found. Outflow parameters are provided in the catalog for the entire sample of 19,508 objects.

\begin{figure*}
    \centering
    \includegraphics[width=\columnwidth]{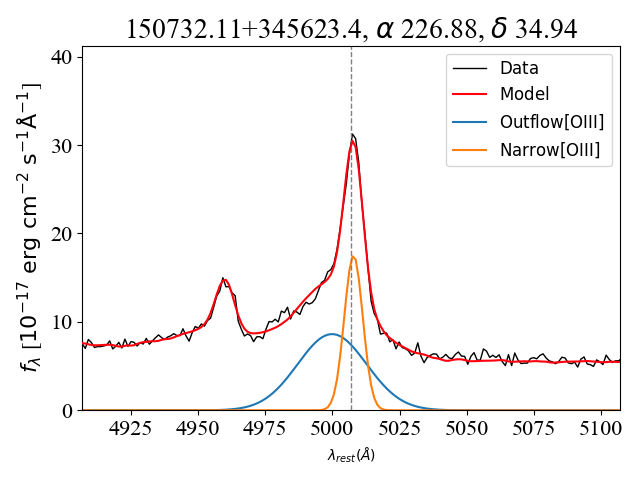}
    \includegraphics[width=\columnwidth]{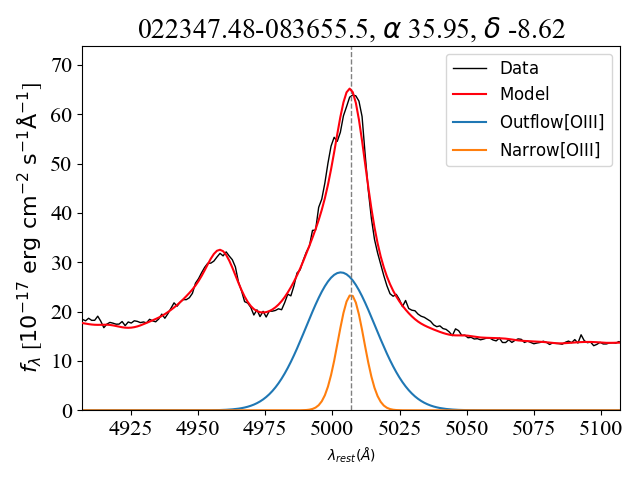}
    \includegraphics[width=\columnwidth]{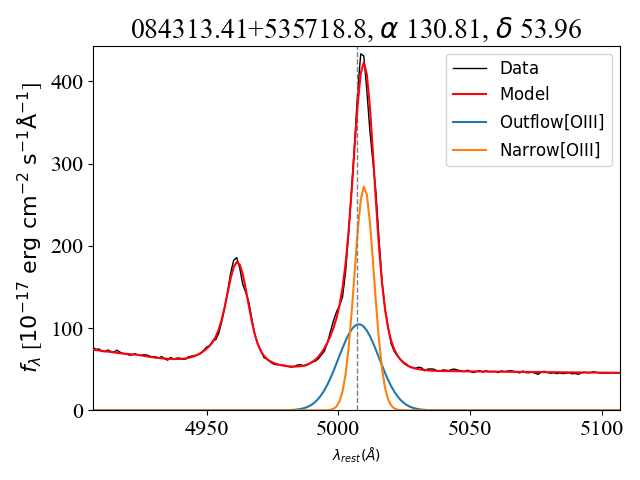}
    \includegraphics[width=\columnwidth]{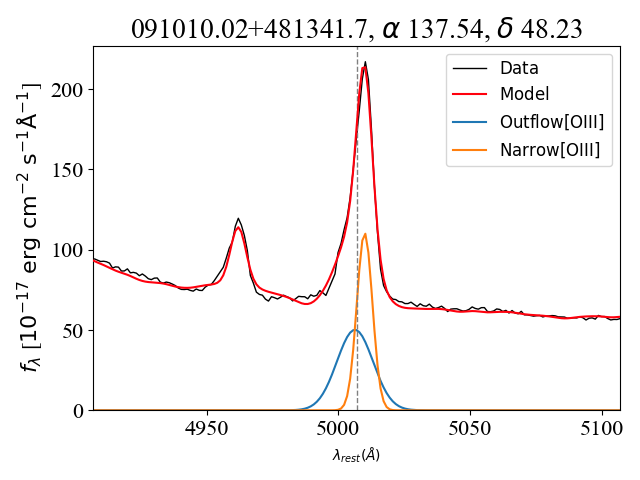}
    \includegraphics[width=\columnwidth]{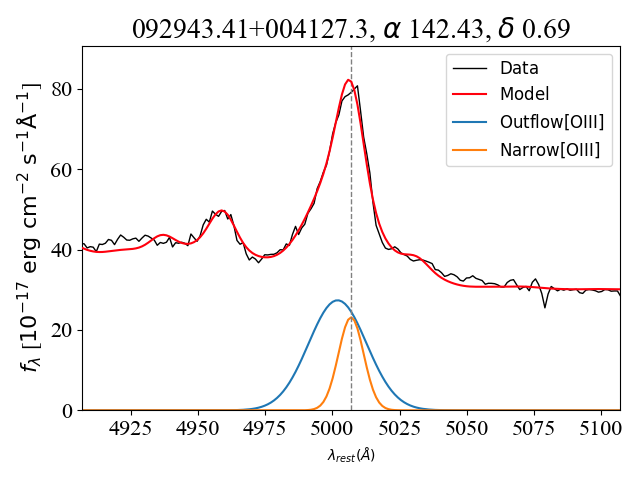}
    \includegraphics[width=\columnwidth]{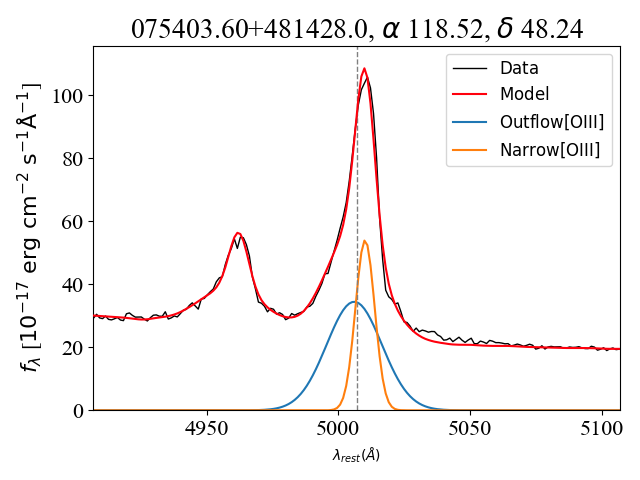}
    
    \caption{Example spectral fits for galaxies which show outflows in the [OIII]~$\lambda$5007 line.}
    \label{fig:spectra}
\end{figure*}

\section{Results}
\subsection{Detection Statistics}

We find that given the sensitivity limit of SDSS, optical coronal line emission is rare in Type~1 quasars, with only 885 ($\sim 4.5$\%) objects out of the sample of 19,508 displaying a detection of at least one coronal line in their SDSS spectra. This detection fraction is higher than that found in the general galaxy population surveyed in the CLASS survey ($\sim 0.03$\%) as would be expected since it includes galaxies that do not host AGNs, but is still surprisingly low given that all objects in the present sample are broad line quasars, where a strong source high energy photons is clearly present. As can be seen from Table \ref{tab:detect_stats}, the detection statistics vary considerably by line, with the [\ion{Ne}{5}] $\lambda$3426 line being the most commonly detected line, followed by the [\ion{Fe}{7}] $\lambda$6087 line. Note that although this sample consists of powerful broad line AGNs, which are characterized by an extremely hard ionizing radiation field, there are very few CLs of the highest ionization potentials detected. There are only 7 [\ion{Fe}{10}] $\lambda$6374 detections, 1 [\ion{Fe}{11}] $\lambda$7892, and not a single [\ion{S}{12}] $\lambda$7611 detection in the entire sample.

\begin{figure*}
    \centering
    \includegraphics[width=\textwidth]{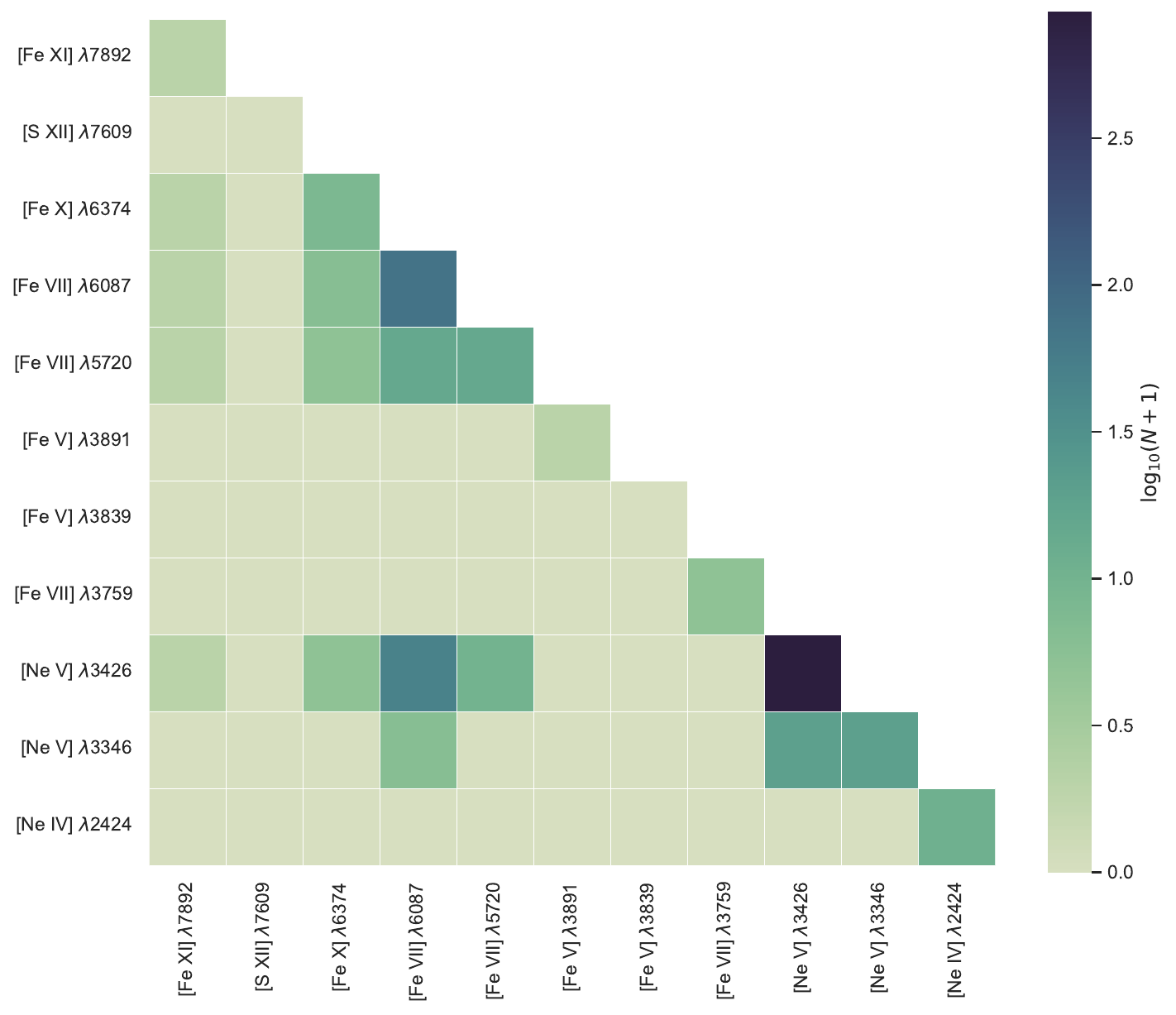}
    \caption{A 2D correlation plot between each pair of lines.  The color of each box represents the number of spectra in which both lines from the horizontal and vertical labels are found together.  Along the diagonal, the color instead represents the pure number of detections of that single line. The color bar is logarithmic, corresponding to $\log_{10}(N+1)$ where $N$ is the number of detections.}
    \label{fig:line_corr}
\end{figure*}

Not only is coronal line emission rare in Type~1 quasars given the sensitivity of SDSS, but it is also clear from this study that it is rare to detect multiple coronal lines in a single spectrum. In Figure \ref{fig:line_corr}, we present a correlation matrix that shows how often lines are detected together in a given spectrum. As can be seen in many cases, only one line is detected. In cases where more than one line is detected, it is most often the [\ion{Ne}{5}] $\lambda$3426 and [\ion{Fe}{7}] $\lambda$6087 lines, which are both found in the spectra of 48 objects. Both lines in the doublet [\ion{Ne}{5}] $\lambda\lambda$3426,3346 are detected in 19 of the objects. 

\subsection{Line Properties}

\begin{figure}[t]
    \centering
    \includegraphics[width=\columnwidth]{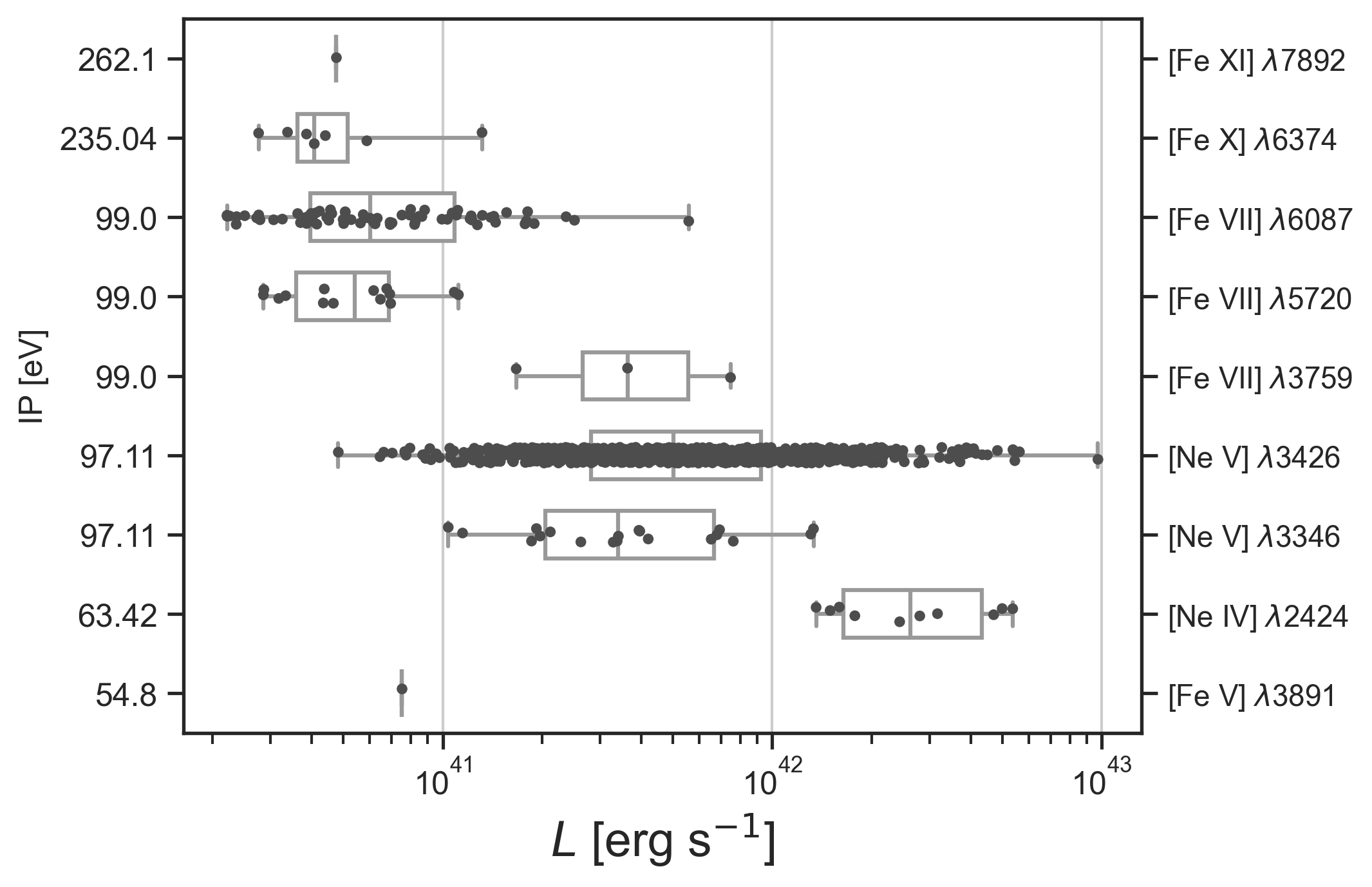}
    \includegraphics[width=\columnwidth]{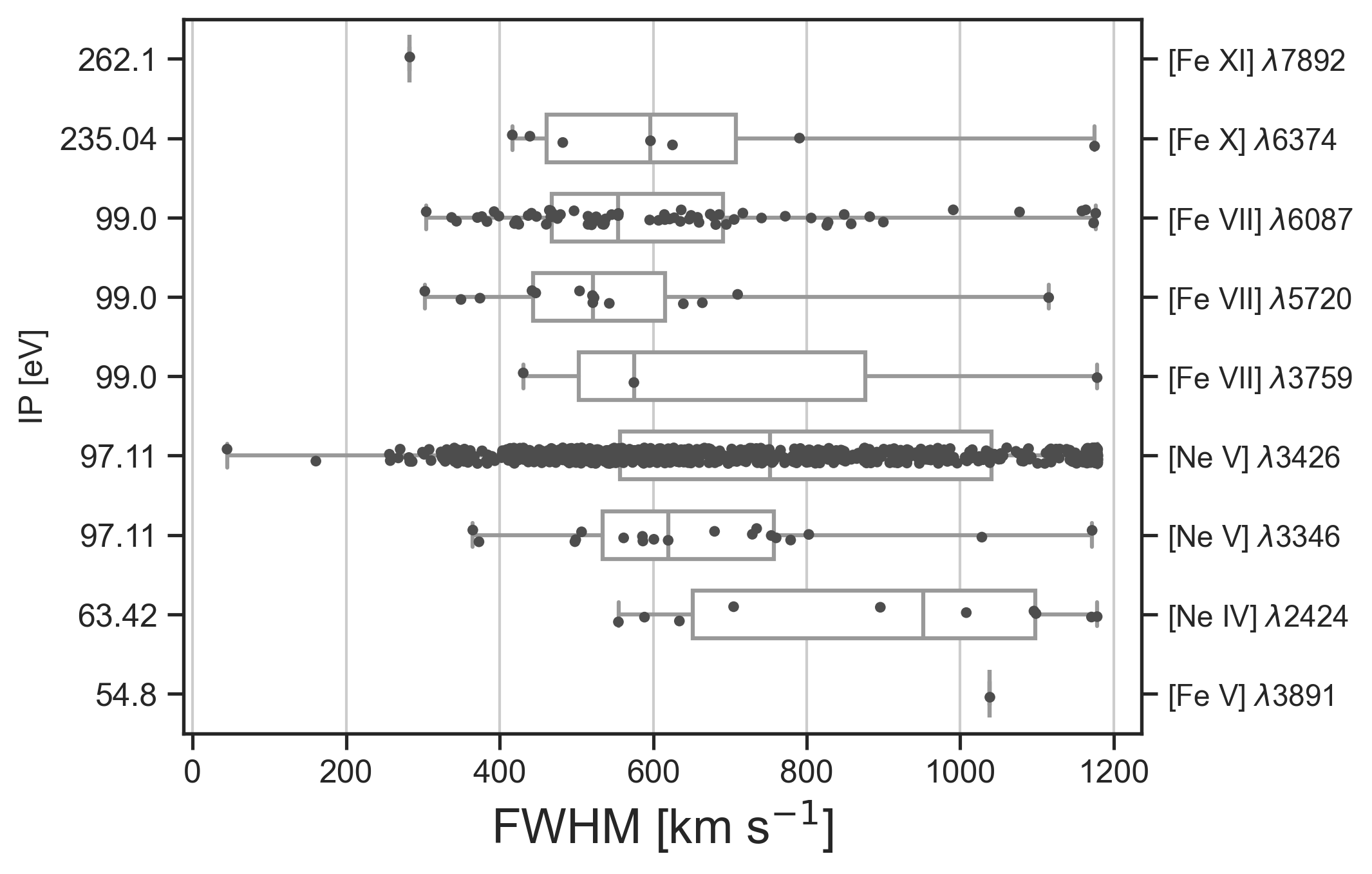}
    \includegraphics[width=\columnwidth]{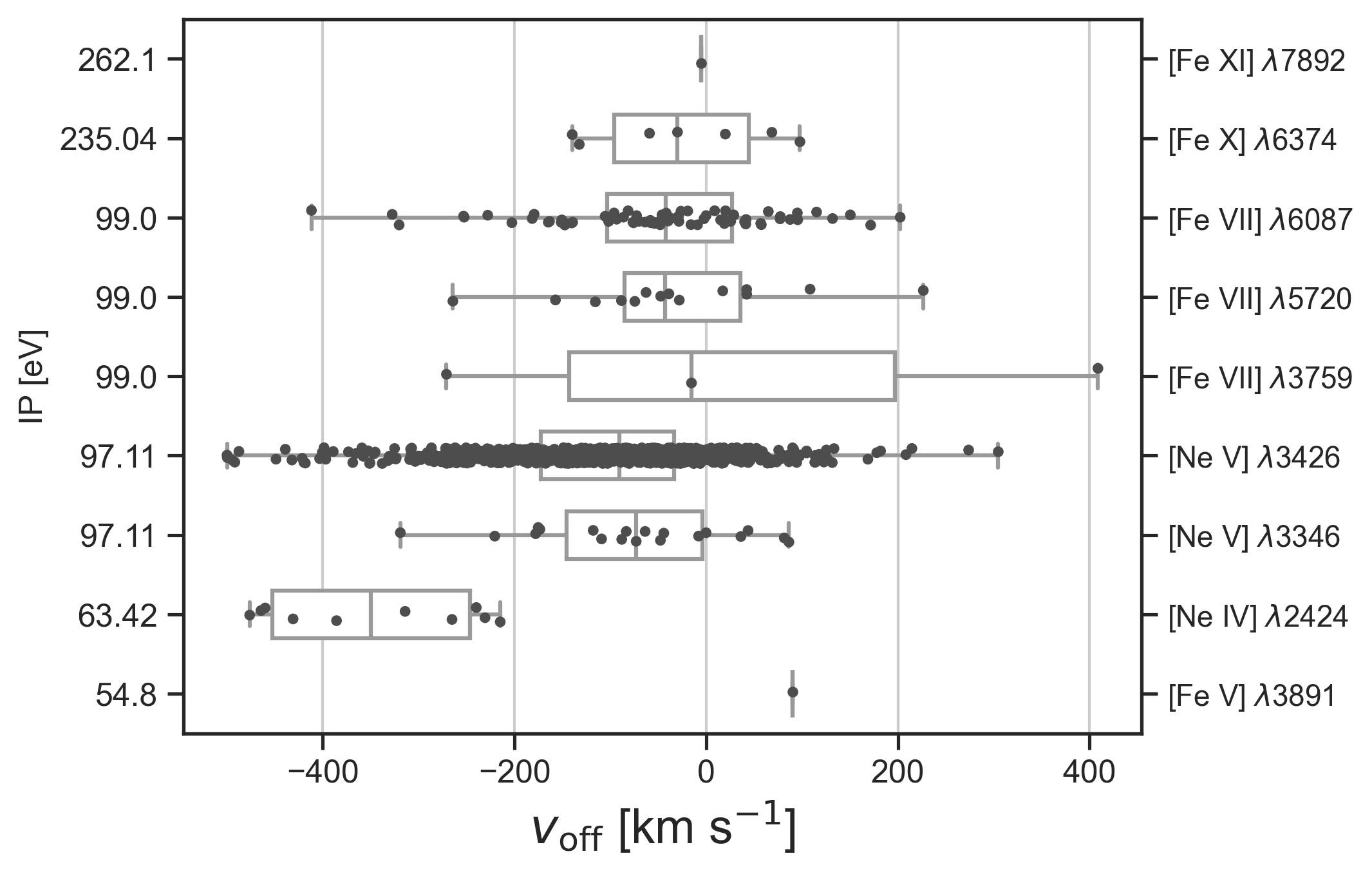}
    \caption{Distributions of each coronal line over the luminosity ($L$), FWHM, and velocity shift (\voff) organized by the IP in eV.  The luminosity is in \ergs, and FWHM and \voff\ are in \kms. Each line is annotated on the right side of the corresponding IP bin. Note that the vertical axes are not scaled linearly; each ionization potential is shown categorically.}
    \label{fig:dist_ip1}
\end{figure}

\begin{figure}[t]
    \centering
    \includegraphics[width=\columnwidth]{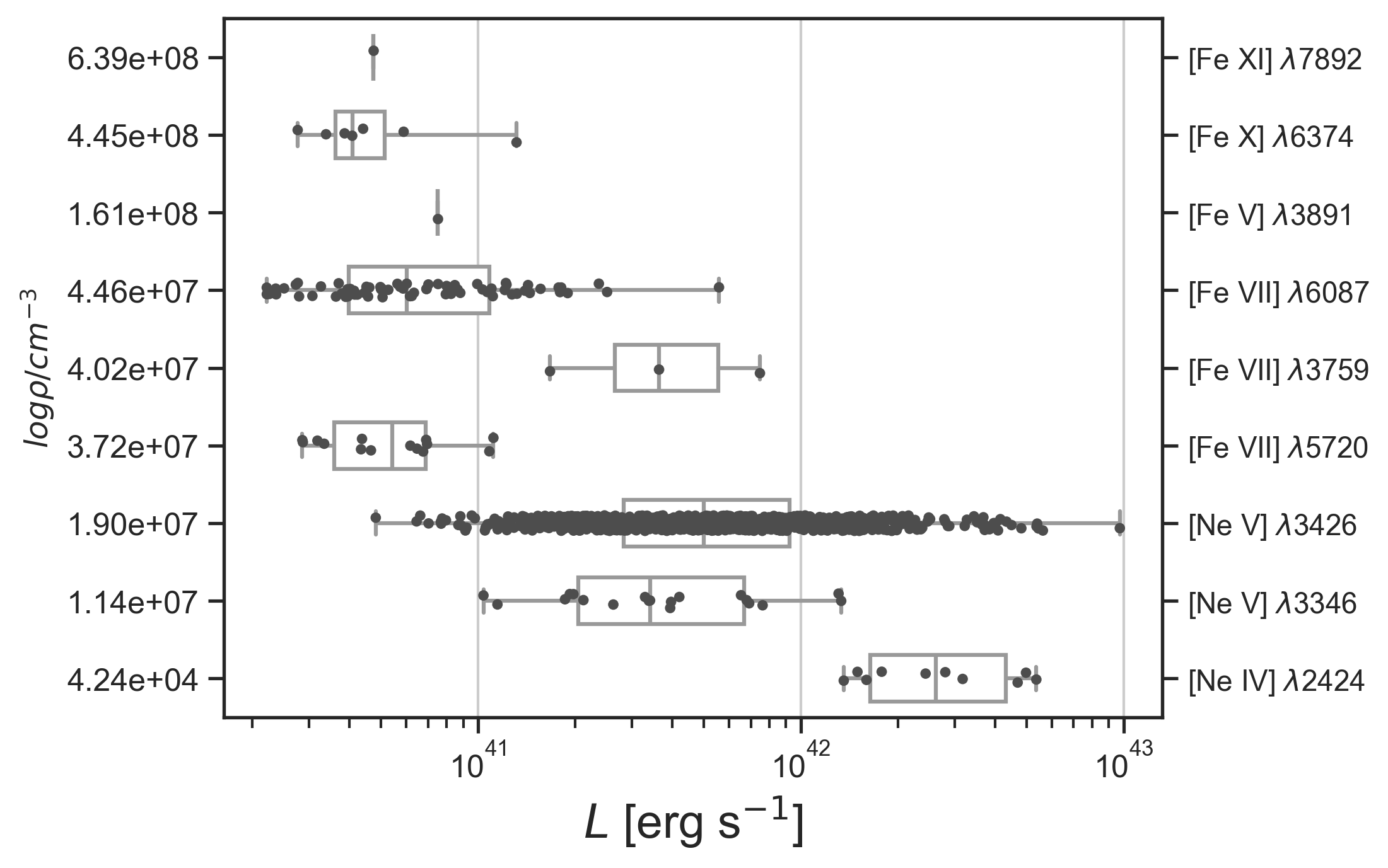}
    \includegraphics[width=\columnwidth]{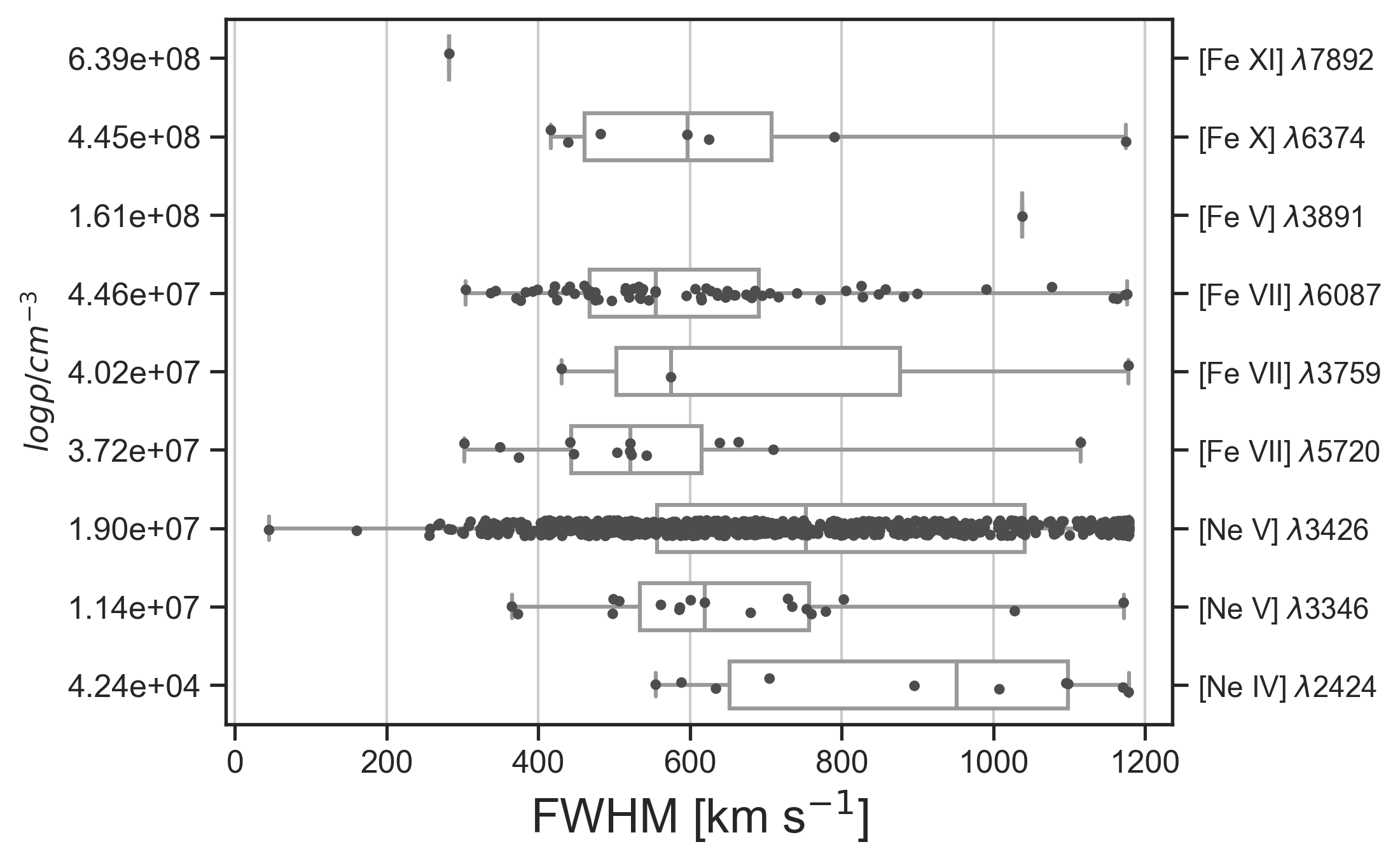}
    \includegraphics[width=\columnwidth]{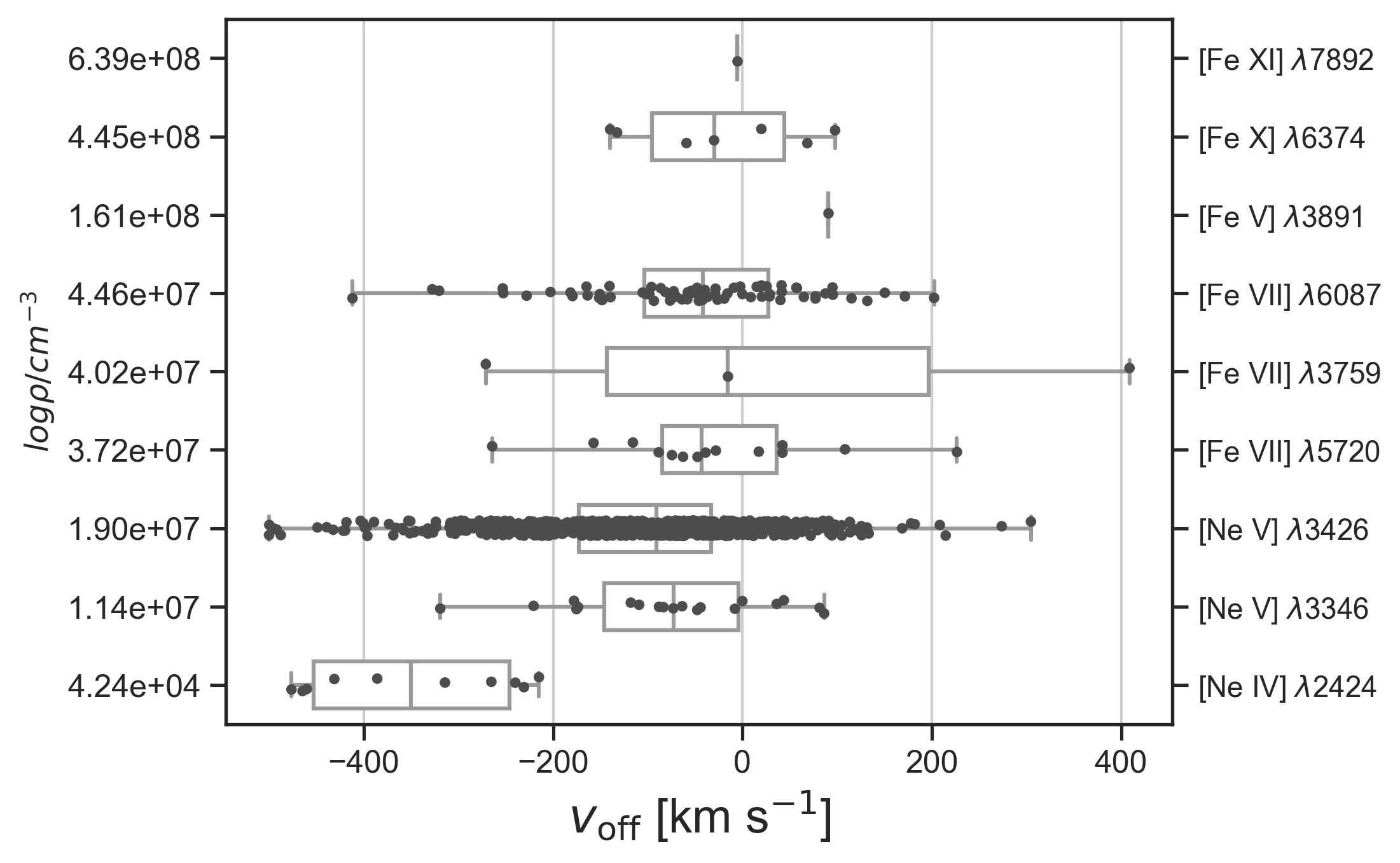}
    \caption{Distributions of each coronal line over the luminosity ($L$), FWHM, and velocity shift (\voff) organized by the critical density in cm$^{-3}$.  The luminosity is in \ergs, and FWHM and \voff\ are in \kms. Each line is annotated on the right side of the corresponding $\log\rho$ bin. Note that the vertical axes are not scaled logarithmically; each critical density is shown categorically.}
    \label{fig:dist_cd1}
\end{figure}

\begin{figure}[t]
    \centering
    \includegraphics[width=\columnwidth]{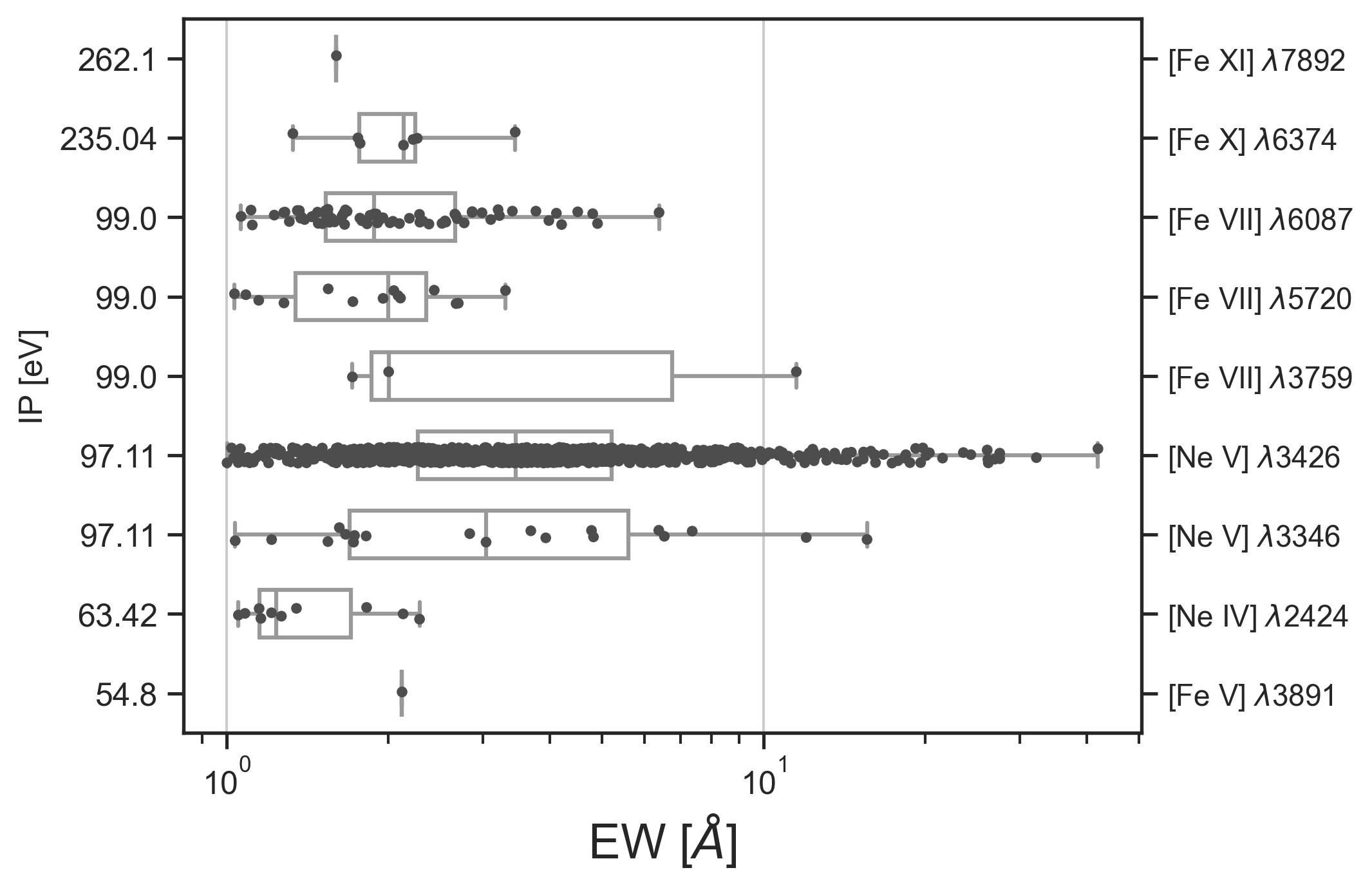}
    \includegraphics[width=\columnwidth]{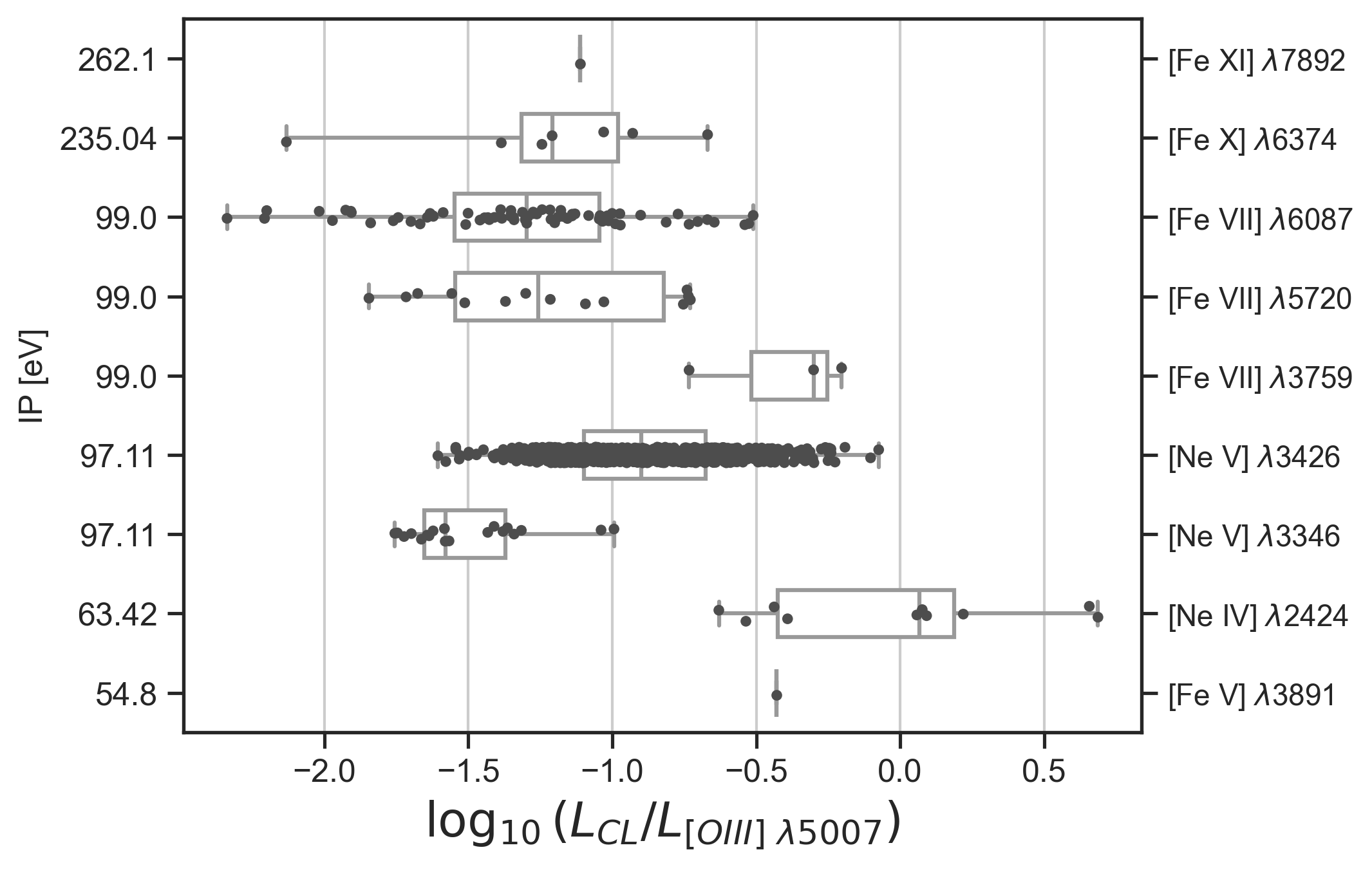}
    \caption{Distributions of each coronal line over the equivalent width (EW) and CL to [\ion{O}{3}] $\lambda$5007 luminosity ratio, organized by the IP in eV. The EW is in \AA. Each line is annotated on the right side of the corresponding IP bin. Note that the vertical axes are not scaled linearly; each ionization potential is shown categorically.}
    \label{fig:dist_ip2}
\end{figure}

\begin{figure}[t]
    \centering
    \includegraphics[width=\columnwidth]{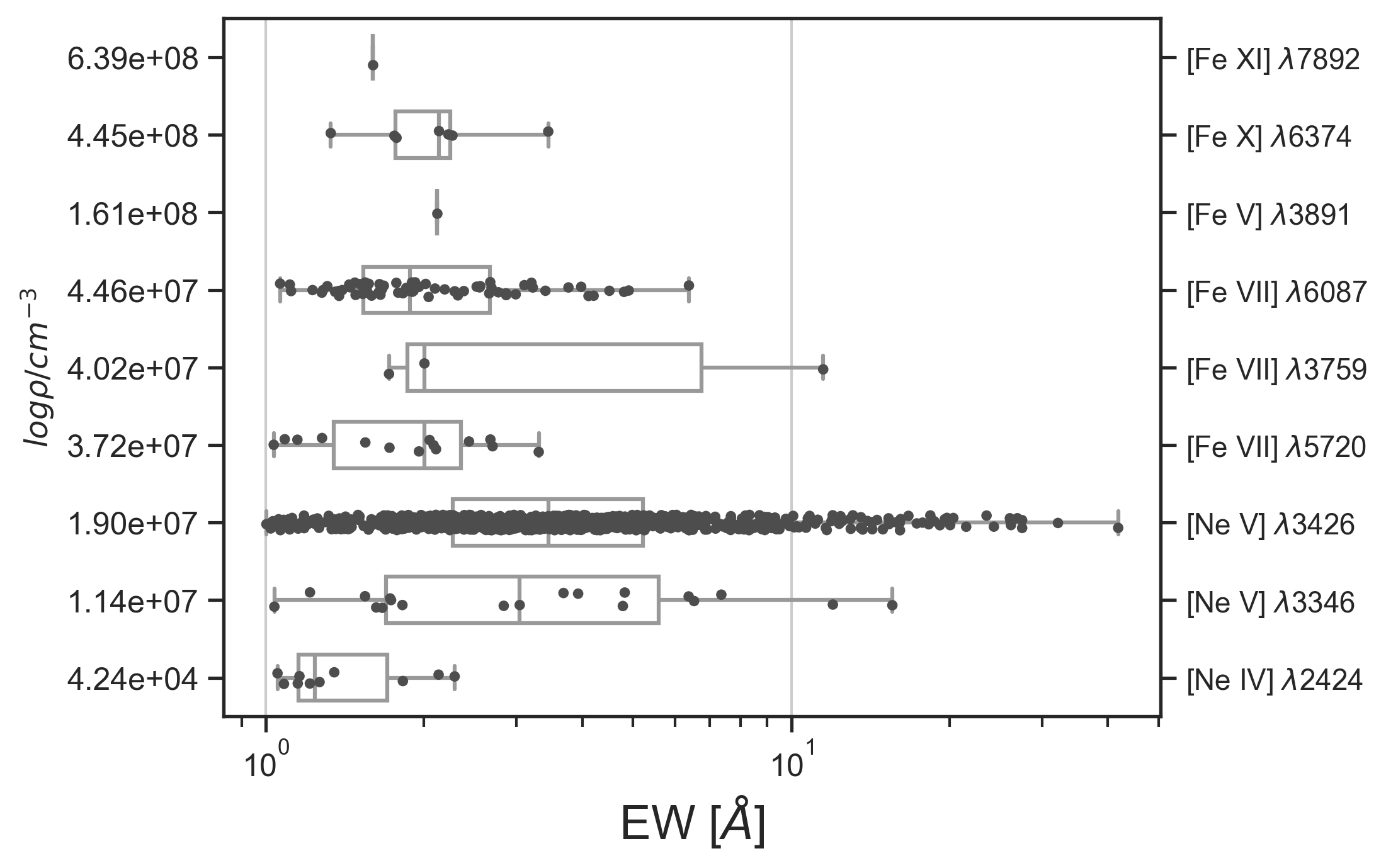}
    \includegraphics[width=\columnwidth]{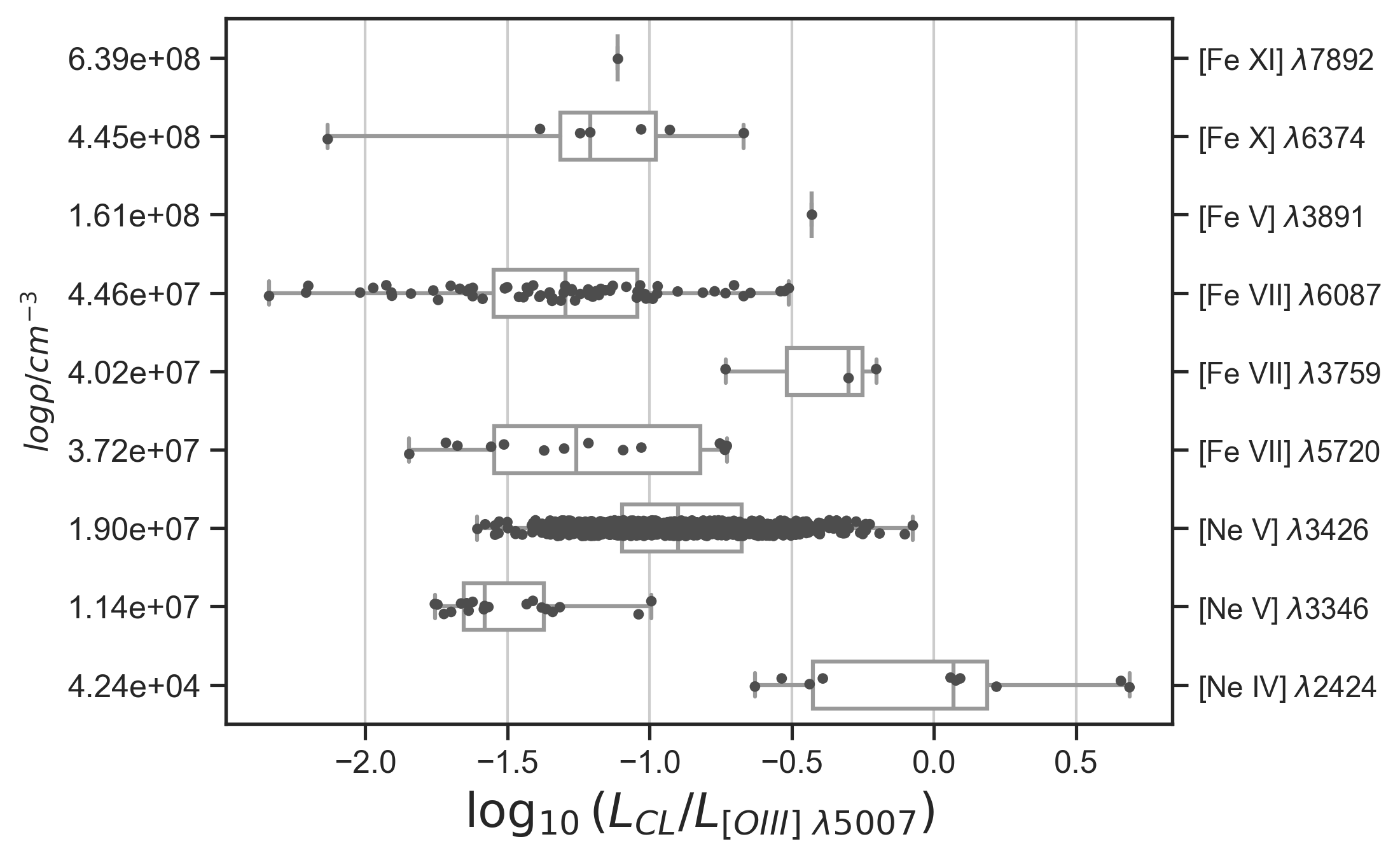}
    \caption{Distributions of each coronal line over the equivalent width (EW) and CL to [\ion{O}{3}] $\lambda$5007 luminosity ratio, organized by the critical density in cm$^{-3}$. The EW is in \AA. Each line is annotated on the right side of the corresponding $\log\rho$ bin. Note that the vertical axes are not scaled logarithmically; each critical density is shown categorically.}
    \label{fig:dist_cd2}
\end{figure}

As can be seen from Table \ref{tab:detect_stats}, the average CL luminosities range from $\sim 5\times10^{40}$--$3\times10^{42}$ \ergs, with average FWHMs extending up to $\sim 1000$~km/s, and average \voff\, typically $< 0$ \kms.  The average luminosities and average FWHMs are somewhat higher than found in the CLASS galaxy survey \citep{2023ApJS..265...21R}, as might be expected given that this sample consists exclusively of powerful quasars. We show the distributions of line properties in our sample in Figures \ref{fig:dist_ip1}, \ref{fig:dist_cd1}, \ref{fig:dist_ip2}, \ref{fig:dist_cd2}, organized vertically by the IP and critical density ($\rho_{\rm crit}$), respectively. As can be seen, there is considerably large dynamic range in the line properties in the sample. The [\ion{Ne}{5}] $\lambda$3426 line displays the largest luminosities extending up to $\sim 10^{43}$ \ergs with a large spread in luminosities in the sample. The range of [\ion{Ne}{5}] $\lambda$3426 luminosities in this sample is consistent with that found in smaller samples of quasars and local Seyferts and X-ray identified AGNs \citep[e.g.,][]{2010A&A...519A..92G, 2022arXiv220906247C}. However, the  [\ion{Ne}{5}] $\lambda$3426 luminosities are in general higher, in some cases by as much as five orders of magnitude, than the luminosities found in the star forming galaxies in the CLASS galaxy survey \citep{2023ApJS..265...21R} and other star forming galaxies in which [\ion{Ne}{5}] $\lambda$3426 emission is detected but is not accompanied by any other signs of nuclear activity \citep[e.g.,][]{2021MNRAS.508.2556I}.  The highest line equivalent widths are found for the [\ion{Ne}{5}] $\lambda$3426 detections, as expected given its prominence. There is no clear trend in line luminosity or equivalent width with ionization potential or critical density; however, it is of note that the few coronal lines detected with the highest ionization potentials and critical densities are at the lower end of the luminosity distribution. 

All CLs show a large spread in FWHM, with many showing broader profiles than is typically seen in the lower ionization potential standard narrow lines. The line widths can extend up to $\sim 1200$~km/s, and display a similar range to what has been observed in other AGN studies \citep[e.g.,][]{1968ApJ...151..807O,1984MNRAS.211P..33P, 1988AJ.....95...45A, 1997A&A...323..707E, 2018ApJ...858...48M}. The velocity offset for the lines also shows a large spread in values in the sample, with more lines displaying a blue shift, in agreement with other coronal line studies in both the optical and infrared \citep[e.g.,][]{2006ApJ...653.1098R,2011ApJ...743..100R, 2011ApJ...739...69M}. In contrast to several other studies on smaller samples of AGNs \cite[e.g.,][]{1997A&A...323..707E, 2006ApJ...653.1098R,2021ApJ...911...70B}, we see no clear trend in FWHM or blueshift with ionization potential or critical density. 

The flux ratios of the various CLs relative to the [\ion{O}{3}] $\lambda$5007 line similarly show a significant spread in values, with the highest values found for the [\ion{Ne}{4}] $\lambda$2424 line, the line corresponding to the lowest ionization potential. The minimum, median, and maximum flux ratios of the detected coronal lines relative to the prominent H$\alpha$ and [\ion{O}{3}] $\lambda$5007 emission lines are shown in Table ~\ref{tab:ratio}. The H$\alpha$ and [\ion{O}{3}] $\lambda$5007 fluxes are calculated from values provided in \cite{shen2011}. The coronal line fluxes can be over two orders of magnitude lower than the H$\alpha$ and [\ion{O}{3}] line fluxes, consistent with what was found in the CLASS galaxy survey \citep{2023ApJS..265...21R}, as well as other coronal line studies  \citep[e.g.,][]{1997A&A...323..707E}. For the vast majority of detections, the CLs are significantly weaker than the H$\alpha$ and [\ion{O}{3}] $\lambda$5007 lines.

\begin{table}[ht]
\caption{Minimum, median, and maximum flux ratios between each coronal line and [O III]$\lambda$5007 and H$\alpha$, in log units}
\begin{center}
\begin{tabular}{c|ccc|ccc}
\hline
	& \multicolumn{3}{c|}{{[}O III{]}} & \multicolumn{3}{c}{H$\alpha$} \\
Line & Min & Median & Max & Min & Median & Max \\
\hline
	\lbrack \ion{Fe}{11}\rbrack$\lambda$7891 & -1.11 & -1.11 & -1.11 & -2.51 & -2.51 & -2.51 \\
	\lbrack \ion{Fe}{10}\rbrack$\lambda$6374 & -2.13 & -1.21 & -0.67 & -2.46 & -2.16 & -1.99 \\
	\lbrack \ion{Fe}{7}\rbrack$\lambda$6087 & -2.34 & -1.30 & -0.51 & -2.54 & -2.24 & -1.60 \\
	\lbrack \ion{Fe}{7}\rbrack$\lambda$5720 & -1.85 & -1.26 & -0.73 & -2.61 & -2.17 & -1.79 \\
	\lbrack \ion{Fe}{5}\rbrack$\lambda$3891 & -0.43 & -0.43 & -0.43 & -2.00 & -2.00 & -2.00 \\
	\lbrack \ion{Fe}{7}\rbrack$\lambda$3758 & -0.73 & -0.30 & -0.20 & -2.00 & -1.64 & -1.29 \\
	\lbrack \ion{Ne}{5}\rbrack$\lambda$3426 & -1.61 & -0.90 & -0.07 & -2.30 & -1.66 & -1.04 \\
	\lbrack \ion{Ne}{5}\rbrack$\lambda$3345 & -1.76 & -1.58 & -0.99 & -2.25 & -1.90 & -1.75 \\
	\lbrack \ion{Ne}{4}\rbrack$\lambda$2424 & -0.63 & 0.07 & 0.69 & --- & --- & --- \\

\hline
\end{tabular}
\end{center}
\label{tab:ratio}
\end{table}

\subsection{Coronal line, WISE, and AGN Bolometric Luminosities}

\begin{figure*}
    \centering
    \includegraphics[width=\columnwidth]{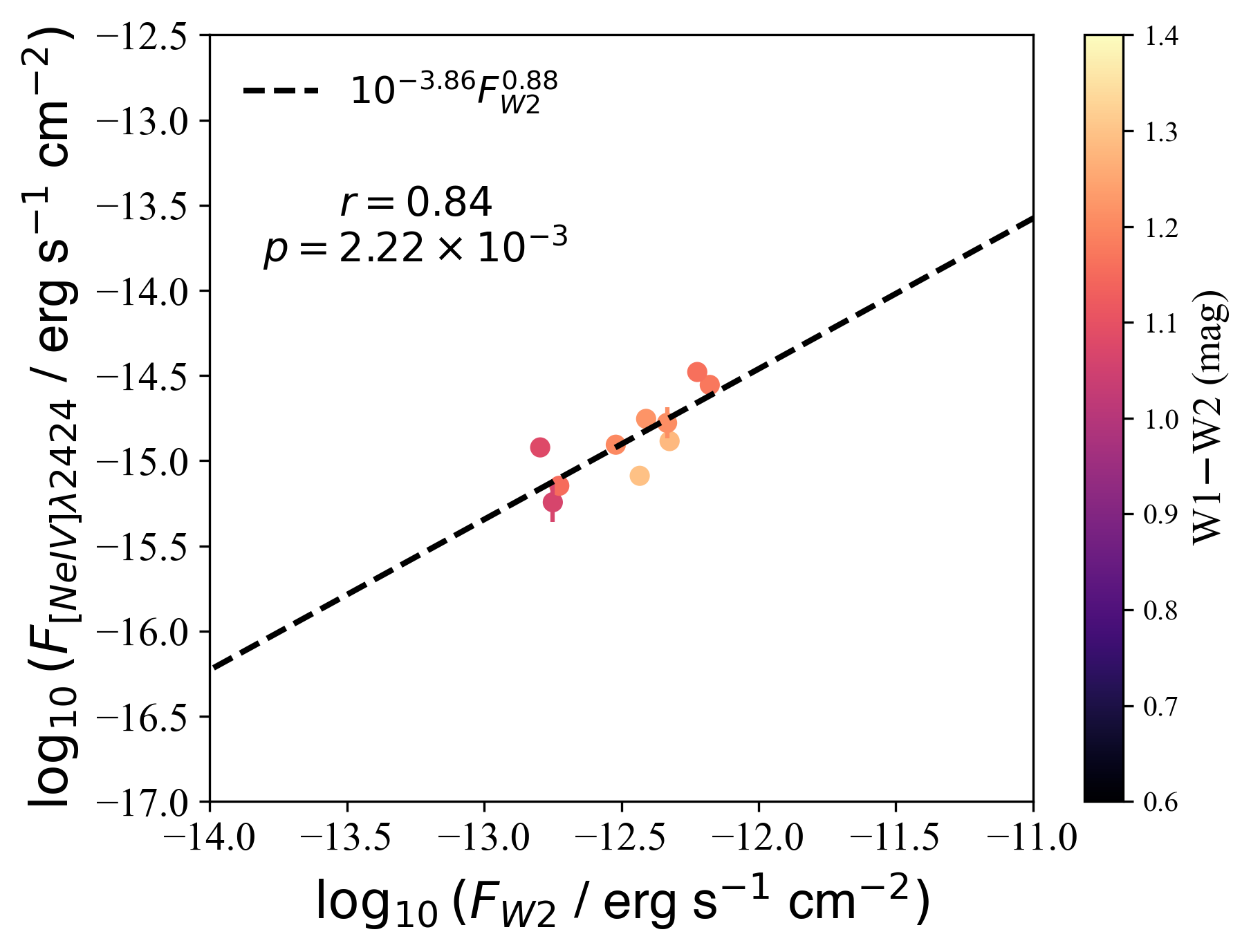}
    \includegraphics[width=\columnwidth]{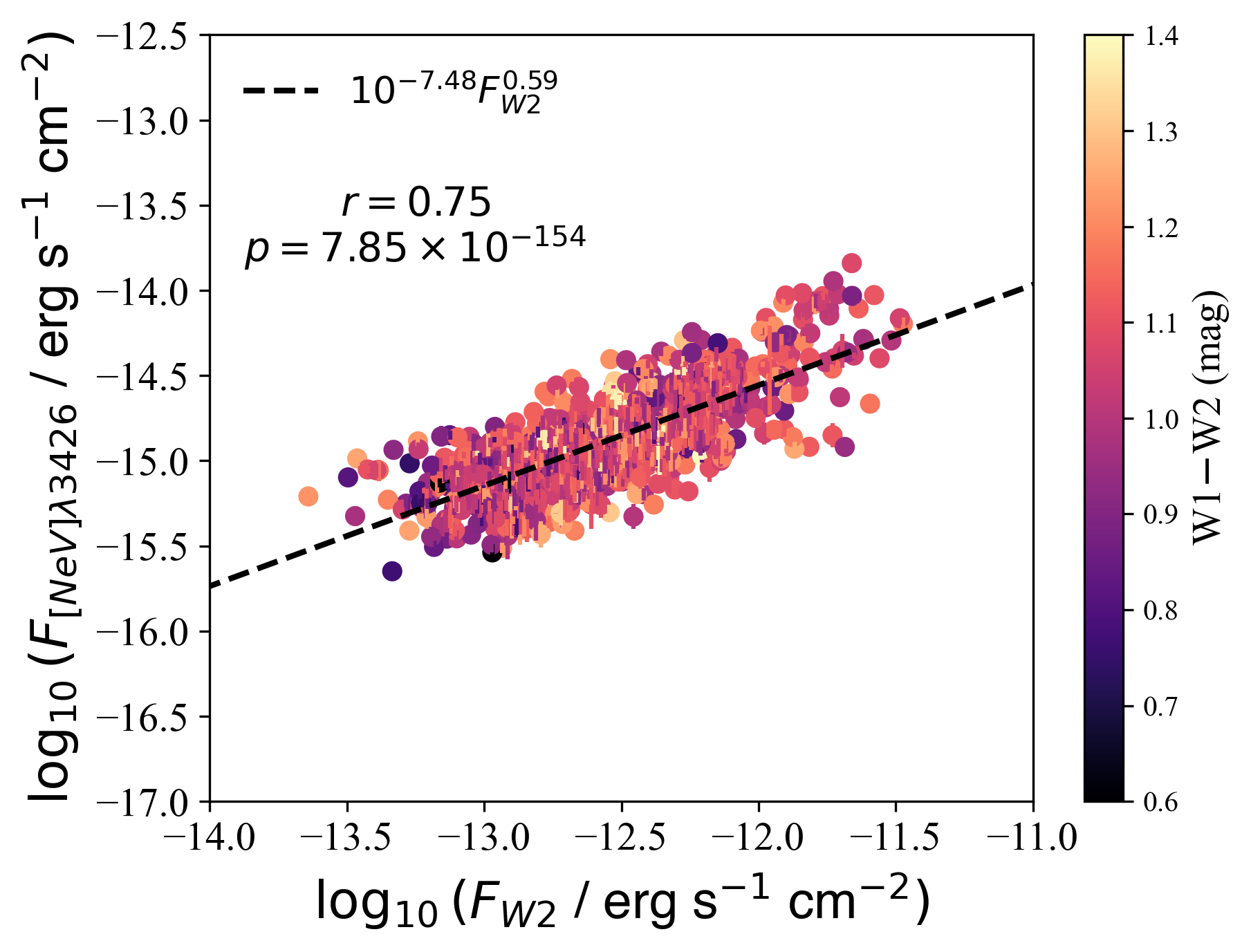}
    \includegraphics[width=\columnwidth]{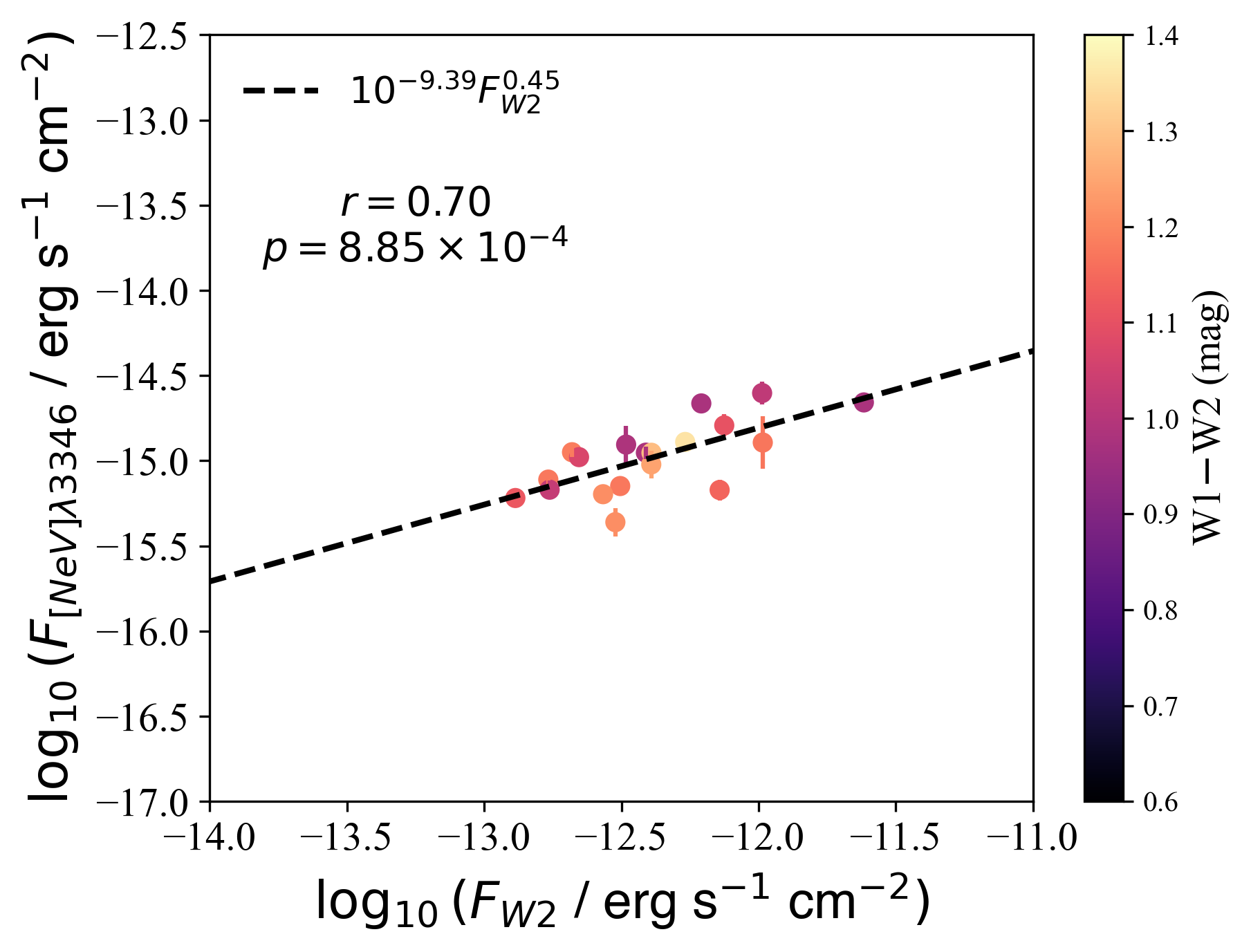}
    \includegraphics[width=\columnwidth]{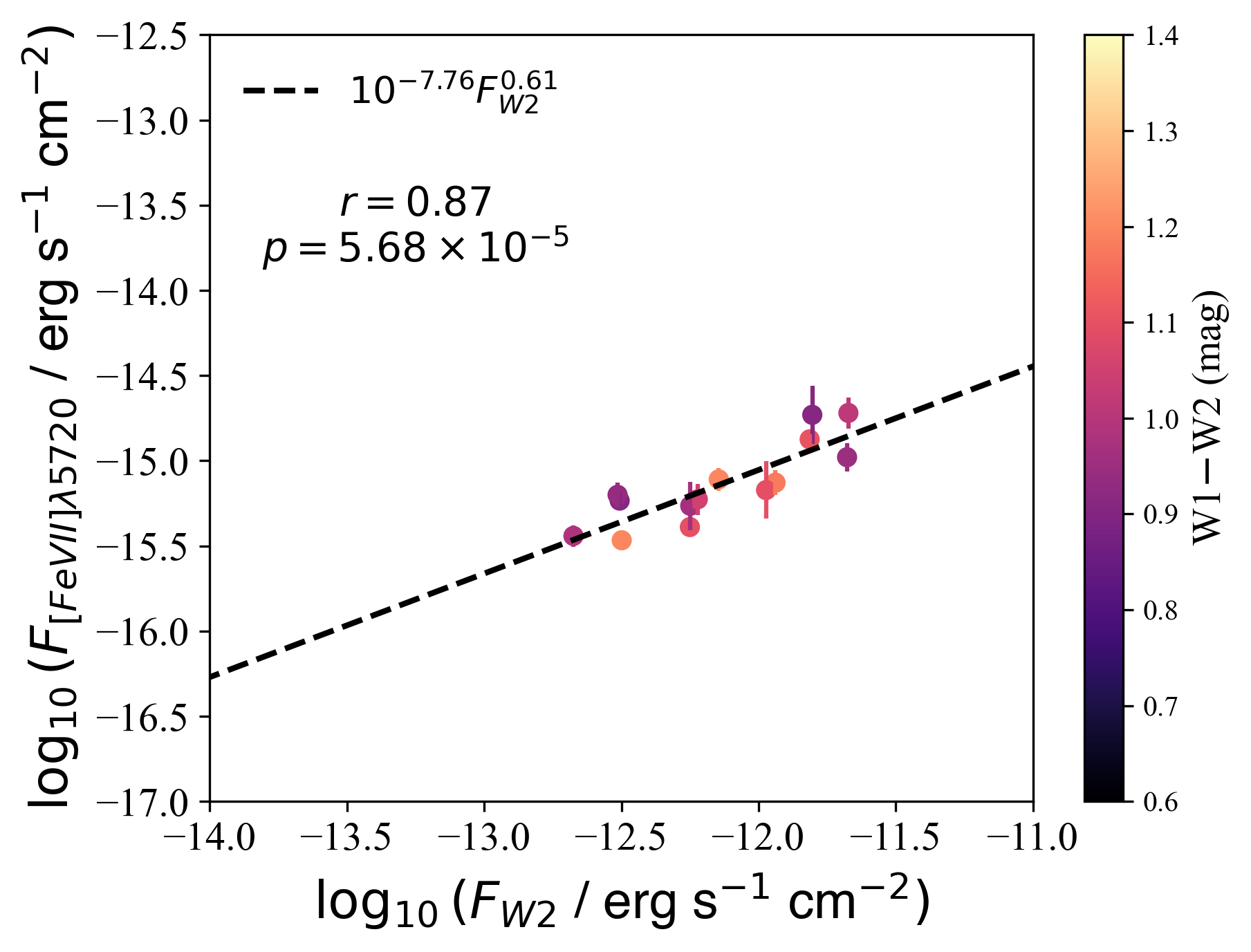}
    \includegraphics[width=\columnwidth]{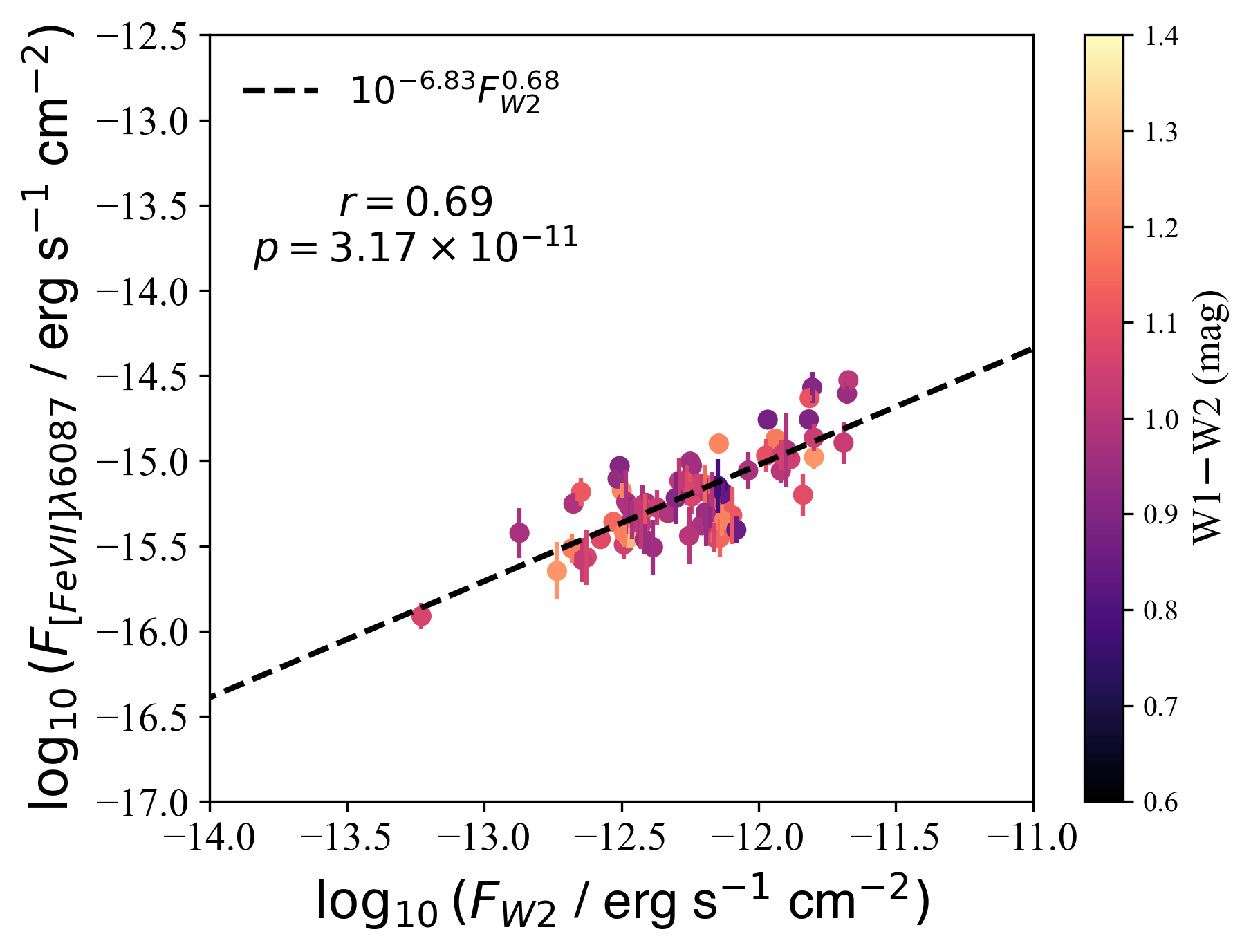}
    \includegraphics[width=\columnwidth]{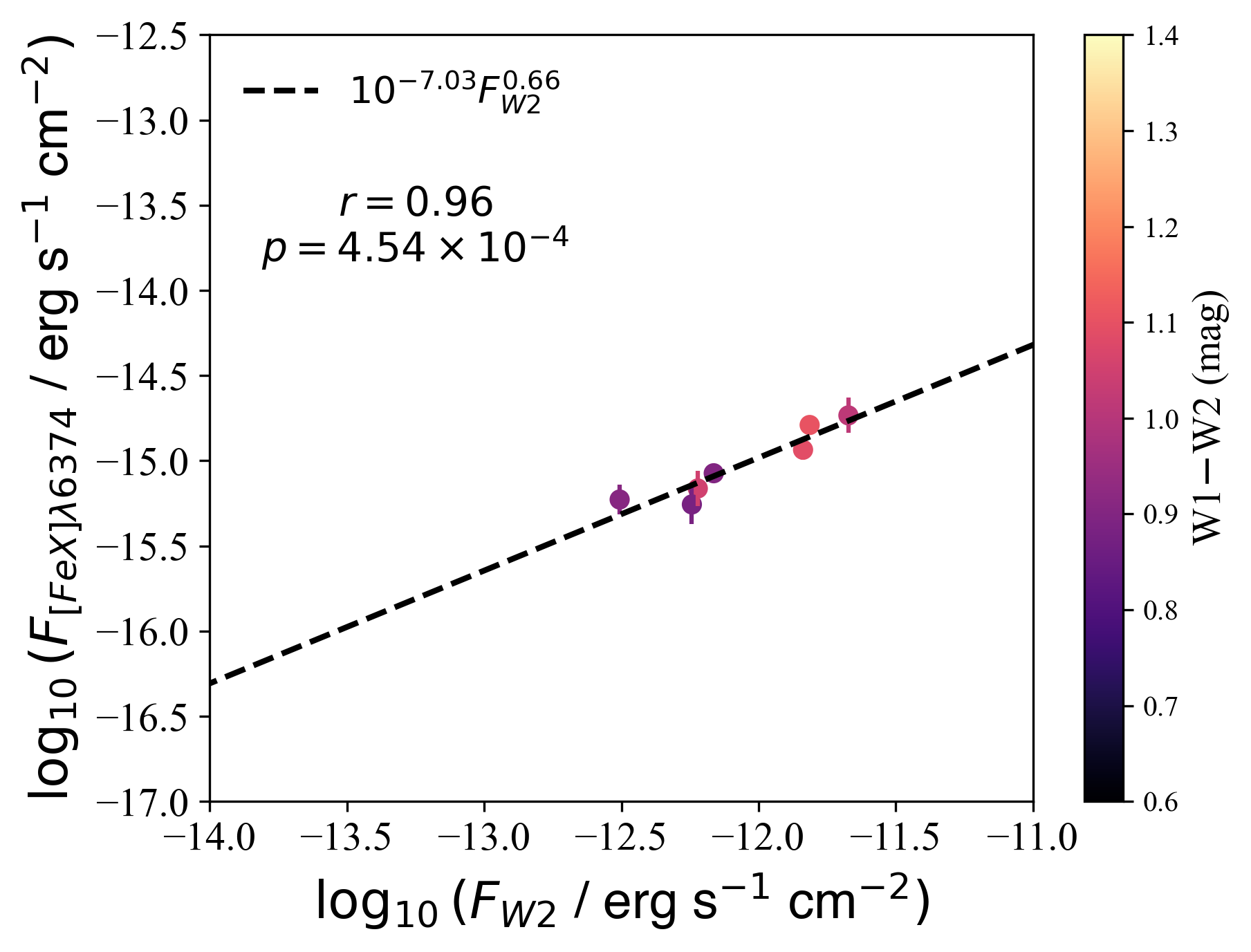}
    \caption{A series of logarithmic scatterplots showing correlations between individual coronal line fluxes and the WISE 2 band fluxes. Each panel has identical vertical and horizontal scales. Best-fit power laws are plotted with dashed lines over the data, and the best-fit parameters are shown in the legend of each plot as a function of $F_{\rm W2}$, the W2 flux. The points are colored according to the W1 $-$ W2 color (in magnitudes). Each panel lists the Spearman correlation coefficient $r$ and corresponding $p$-value of the distribution.}
    \label{fig:w2_lum}
\end{figure*}

We replicate the trend reported in the CLASS galaxy survey \citep{2023ApJS..265...21R}, in which we find a strong correlation between the coronal line fluxes, and the WISE 4.6 $\mu$m (W2) flux. In Figure  \ref{fig:w2_lum}, we show a selection of scatter plots of coronal line fluxes and 4.6 $\mu$m (W2) flux, with the best-fit power law displayed in the upper left of each panel.  We exclude in these plots those lines for which there are very few detections. These correlations indicate that the W2 flux, which is available with an all-survey, can be used to predict the coronal line strength.

\begin{figure}
    \centering
    \includegraphics[width=0.45\textwidth]{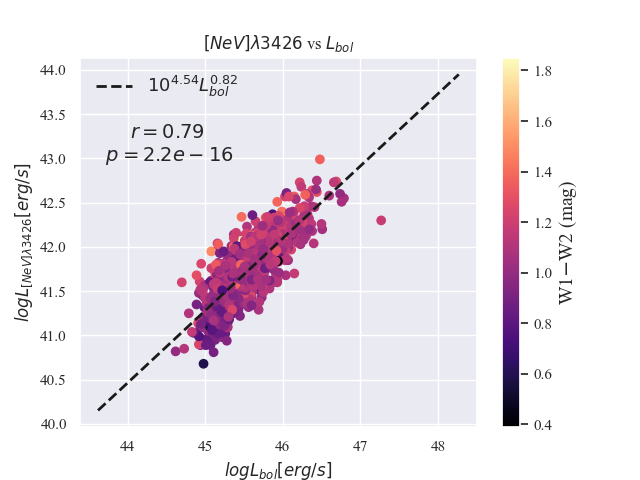}

    \caption{ [\ion{Ne}{5}] $\lambda$3426 line luminosity vs. the bolometric luminosity for all the detections in the sample, with the best-fit power law displayed. The fit parameters and Spearman rank correlation coefficients are shown in the upper left corner.}
    \label{fig:nev_lum}
\end{figure}

Since estimates of the bolometric luminosity of the AGN are available for the sample, we explore the relationship between the line strength and the bolometric luminosity for the most prominent line. In Figure~\ref{fig:nev_lum}, we plot the [\ion{Ne}{5}] $\lambda$3426 line luminosity vs. the bolometric luminosity for all the detections in the sample. Bolometric luminosities listed are taken directly from \citet{shen2011} and are based on the 5100~\AA\  continuum luminosity. We note that the 5100~\AA\  continuum luminosity can be contaminated by the host galaxy for low redshift galaxies, and the bolometric corrections are highly uncertain, as emphasized by the cautionary notes listed in \cite{shen2011}. As can be seen, there is a significant correlation for detections, indicating that the coronal line luminosity is a good indicator of the AGN bolometric luminosity for coronal line emitting AGNs. Coronal line flux correlates with the X-ray flux, which is a good indicator of the AGN bolometric luminosity for non Compton-thick AGNs, in smaller samples of AGN reported in the literature also suggest that the coronal line luminosity can be used to estimate the bolometric luminosity of the AGN \cite[e.g.,][]{10.1111/j.1365-2966.2009.14961.x, 2010A&A...519A..92G, 2017MNRAS.467..540L}. We note that the slope of the relations are systematically higher than found in the CLASS galaxy sample \citep{2023ApJS..265...21R}, particularly for the [\ion{Ne}{5}] $\lambda$3426  relation.

\subsection{Relationship between coronal line emission and black hole properties}

\begin{figure*}[h]
    \centering
    \includegraphics[width=\columnwidth]{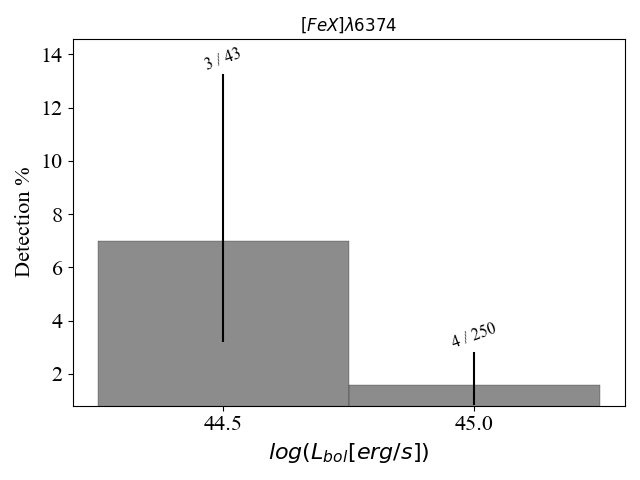}
    \includegraphics[width=\columnwidth]{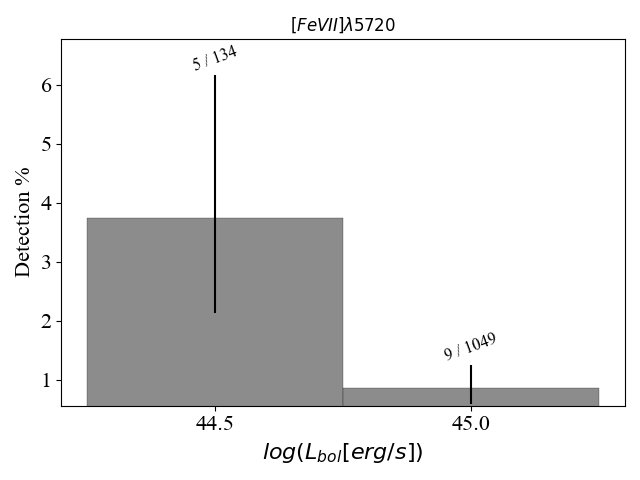}
    \includegraphics[width=\columnwidth]{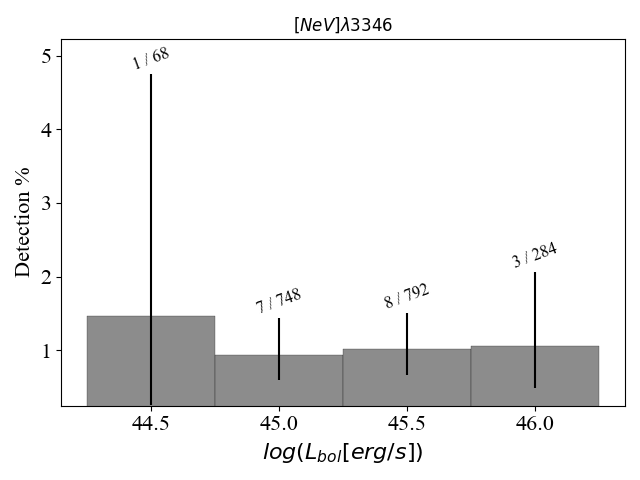}
    \includegraphics[width=\columnwidth]{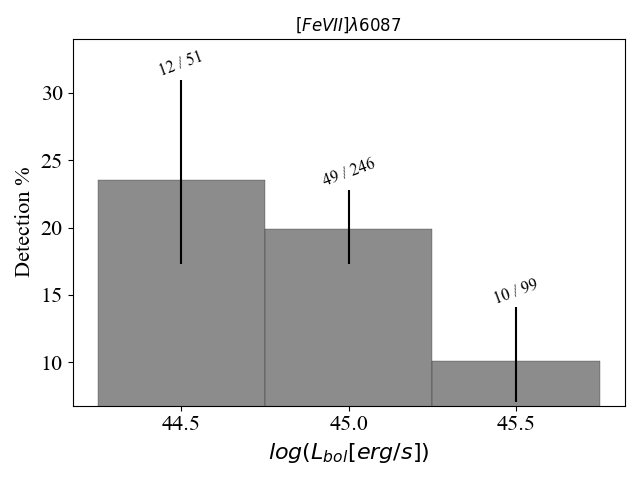}
    \includegraphics[width=\columnwidth]{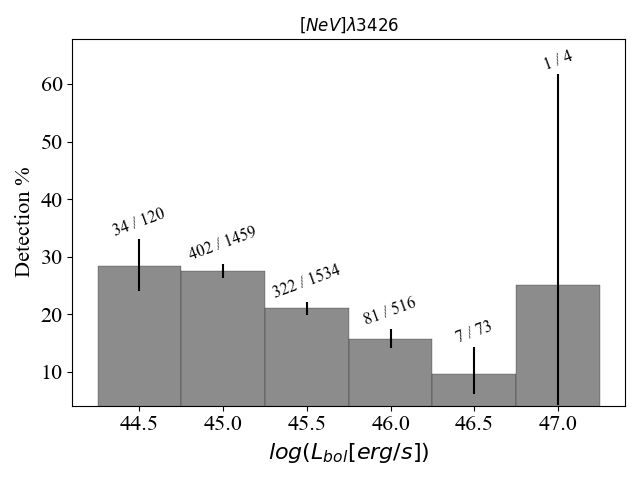}
    \includegraphics[width=\columnwidth]{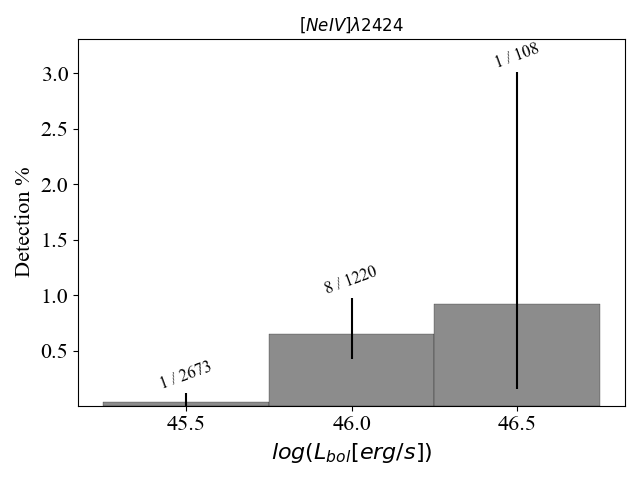}
    \caption{Histograms of detection fractions for the various coronal lines indicated as a function of bolometric luminosity. Number counts are listed for each bin and error bars are calculated using binomial statistics using a 68 percent confidence interval. }
    \label{fig:detect_histo_lum}
\end{figure*}

\begin{figure*}[h]
    \centering
    \includegraphics[width=\columnwidth]{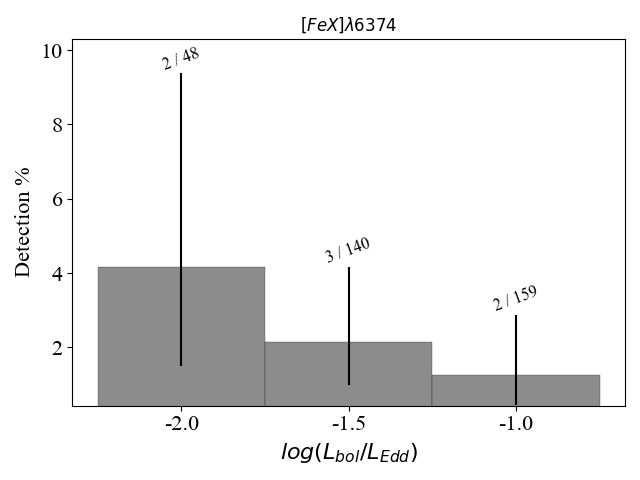}
    \includegraphics[width=\columnwidth]{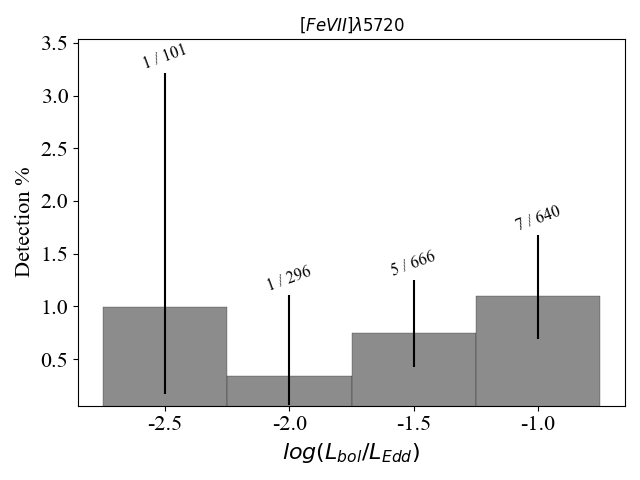}
    \includegraphics[width=\columnwidth]{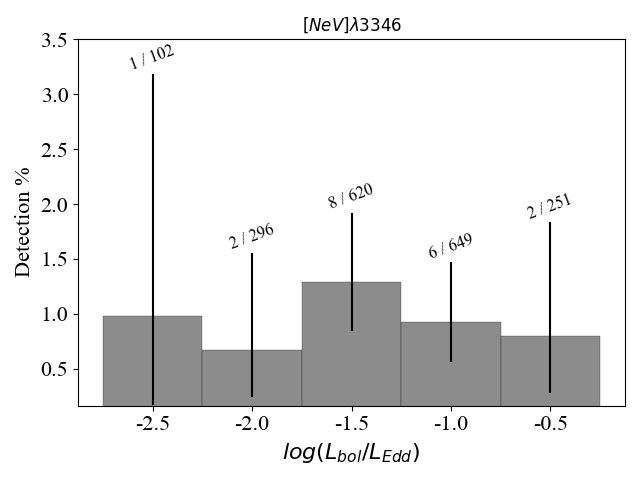}
    \includegraphics[width=\columnwidth]{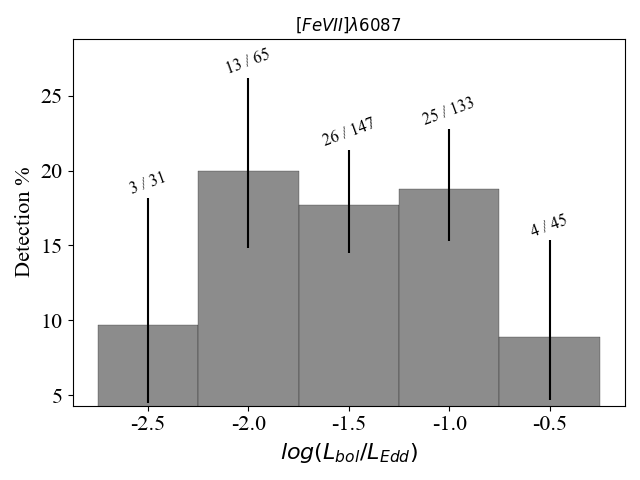}
    \includegraphics[width=\columnwidth]{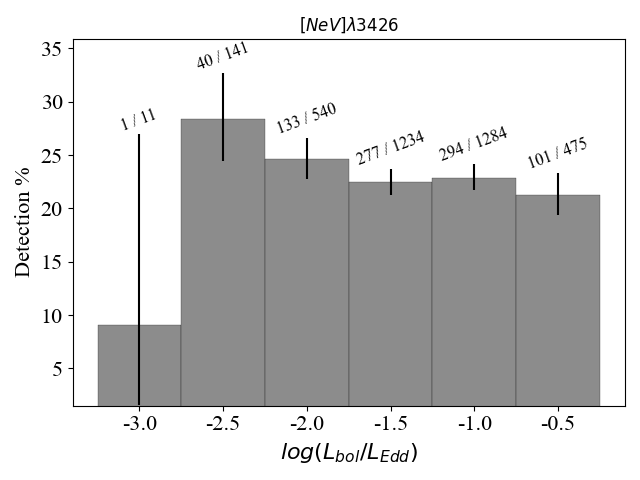}
    \includegraphics[width=\columnwidth]{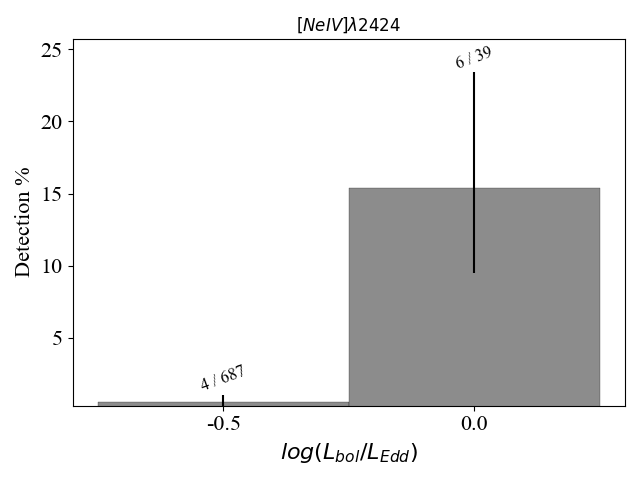}
    \caption{Histograms of detection fractions for the various coronal lines indicated as a function of Eddington ratio. Number counts are listed for each bin and error bars are calculated using binomial statistics using a 68 percent confidence interval. }
    \label{fig:detect_histo_edd}
\end{figure*}

\begin{figure*}[h]
    \centering
    \includegraphics[width=\columnwidth]{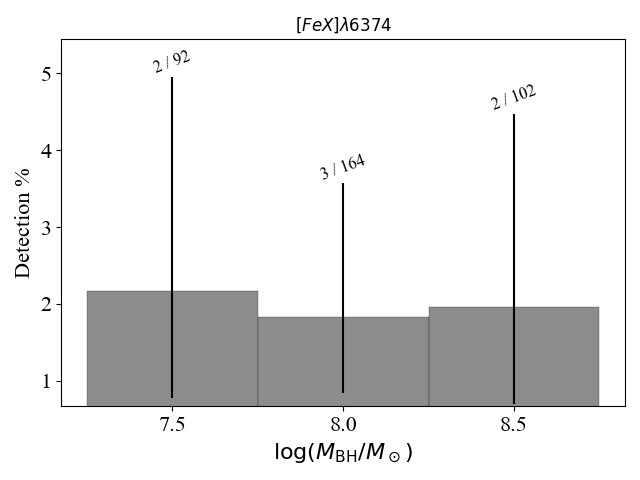}
    \includegraphics[width=\columnwidth]{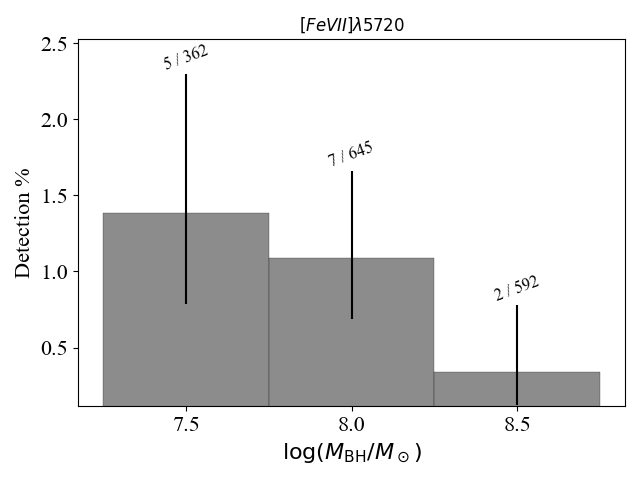}
    \includegraphics[width=\columnwidth]{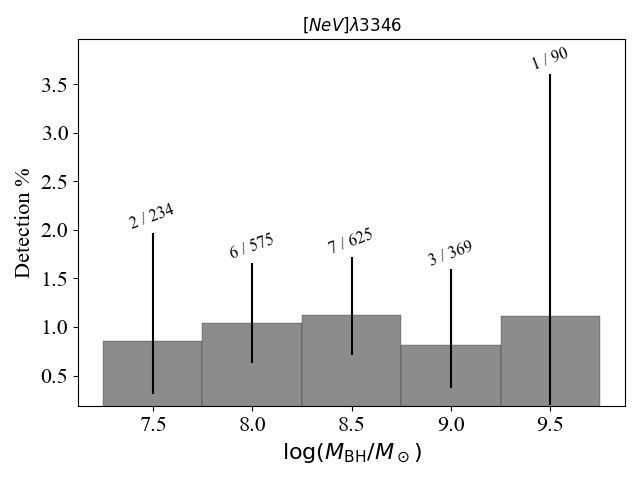}
    \includegraphics[width=\columnwidth]{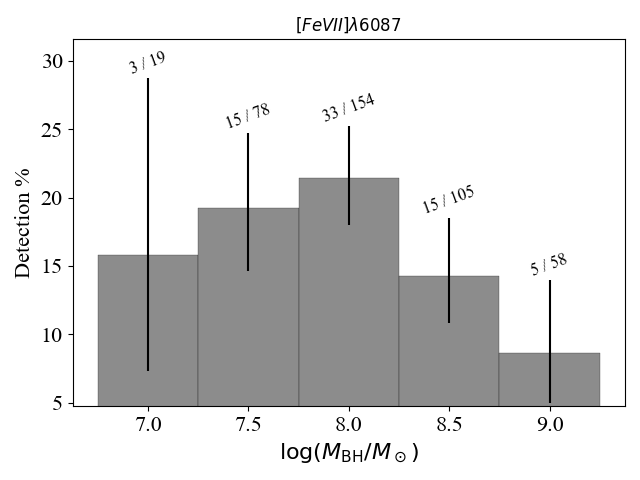}
    \includegraphics[width=\columnwidth]{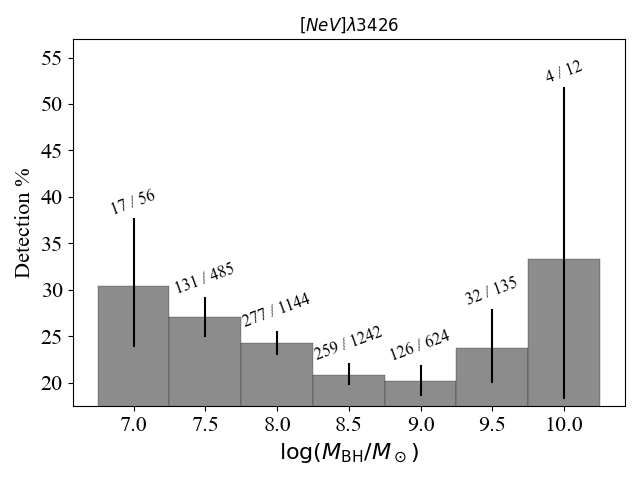}
    \includegraphics[width=\columnwidth]{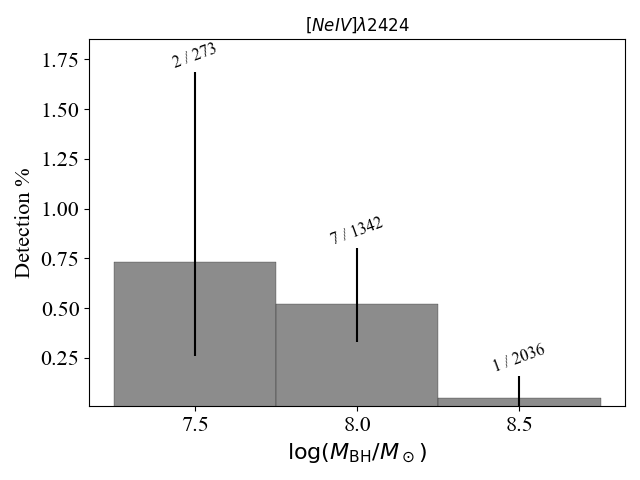}
    \caption{Histograms of detection fractions for the various coronal lines indicated as a function of mass. Number counts are listed for each bin and error bars are calculated using binomial statistics using a 68 percent confidence interval. }
    \label{fig:detect_histo_mass}
\end{figure*}

In this section, we explore whether the CL properties depend on black hole or accretion properties in the sample. If the CLs are produced by photoionized gas irradiated by the AGN, line flux ratios might depend on the AGN properties, as might be expected given that the AGN radiation field may vary with black hole mass and Eddington ratio. Because of the rarity of detections, and the fact that there are very few galaxies in which multiple coronal lines are detected, we cannot easily explore whether such a dependence is seen in our sample. However, we explore the demographics of the coronal line emitters in our sample to see whether the detection fraction depends on black hole mass or Eddington ratio. In Figures~\ref{fig:detect_histo_lum}, ~\ref{fig:detect_histo_edd}, and ~\ref{fig:detect_histo_mass}, we show the detection fraction as a function of AGN properties for several coronal lines. We indicate the number counts in each bin, and error bars using binomial statistics using a two-sided 68 percent confidence interval. Because the coronal line luminosities, and therefore detectability, will likely depend on these properties as well as redshift and the sensitivity of each spectrum, we only consider objects in which the upper limit of the given coronal line relative to the [\ion{O}{3}] $\lambda$5007 line flux is below the minimum ratio of all detections for the given line. By so doing, we ensure that we are exploring if the given coronal line is enhanced relative to the [\ion{O}{3}] $\lambda$5007 line flux in AGNs in a specific range of black hole mass, bolometric luminosity, or Eddington ratio, or if there is no apparent trend with AGN property. As can be seen from these figures, the coronal line detections tend to be found in the less luminous objects with lower black hole masses and lower Eddington ratios; however, given the low number of detections, the only trend that is statistically significant is that coronal line emitters are not found in objects with the highest Eddington ratios. 


\subsection{Relationship between coronal line emission and  [OIII]~$\lambda$5007 outflow properties}

\begin{figure*}
    \centering
    \includegraphics[width=\columnwidth]{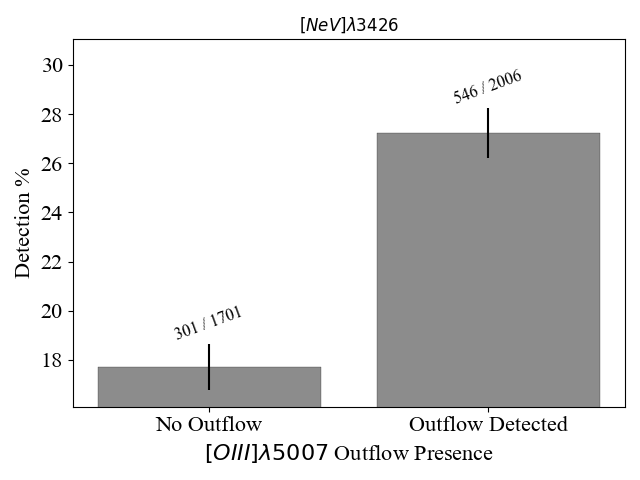}
    \caption{}
    \label{fig:outflow_det_frac}
\end{figure*}

We find that 4,296 out of 19,508 of the galaxies in the sample display an outflow component in the [\ion{O}{3}]~$\lambda$5007 emission line. Of the 885 AGNs in the sample with at least one coronal line detected, we find that 564 ($\approx 64\%$) display an [\ion{O}{3}]~$\lambda$5007 outflow. In Figure ~\ref{fig:outflow_det_frac}, we compare the coronal line detection fractions in galaxies with and without [\ion{O}{3}]~$\lambda$5007 outflows for the [\ion{Ne}{5}] $\lambda$3426 line, for which there are the greatest number of detections.  Uncertainties on the detection fractions are given by binomial counting statistics with a two-sided 68 per cent confidence interval. As can be seen for most CLs, there is a clear trend with higher [\ion{Ne}{5}] $\lambda$3426  detection fractions in AGNs that display outflows in their [\ion{O}{3}]~$\lambda$5007 line profiles.The trend is statistically significant in the [\ion{Ne}{5}] $\lambda$3426 line, for which there are the most numerous detections. In Tables ~\ref{tab:outflow_voff} and ~\ref{tab:outflow_fwhm}, we show the average velocity offset and FWHM of the [OIII]~$\lambda$5007 outflow components for CL detections and non-detections for each CL. Given the limited statistics of the sample, there is no clear trend with offset velocities and outflow FWHM.

\begin{table}[ht]
\caption{[OIII] Outflow Velocity Offset $[km/s]$ average in coronal line detections vs. non-detections}
\begin{center}
\begin{tabular}{c|c|c}
\hline
Line & CL Detections & CL Non-detections \\
\hline
	\lbrack \ion{Fe}{11}\rbrack$\lambda$7891 & $-279.72 \pm 0.00$ & $-106.06 \pm 115.87$ \\
	\lbrack \ion{Fe}{10}\rbrack$\lambda$6374 & $-112.02 \pm 96.08$ & $-123.29 \pm 125.61$ \\
	\lbrack \ion{Fe}{7}\rbrack$\lambda$6087 & $-98.97 \pm 109.22$ & $-125.89 \pm 124.92$ \\
	\lbrack \ion{Fe}{7}\rbrack$\lambda$5720 & $-104.11 \pm 124.30$ & $-126.40 \pm 124.70$ \\
	\lbrack \ion{Ne}{5}\rbrack$\lambda$3426 & $-119.90 \pm 121.86$ & $-124.35 \pm 126.39$ \\
	\lbrack \ion{Ne}{5}\rbrack$\lambda$3345 & $-83.84 \pm 132.03$ & $-121.52 \pm 130.51$ \\
	\lbrack \ion{Ne}{4}\rbrack$\lambda$2424 & $-234.09 \pm 123.17$ & $-115.87 \pm 127.16$ \\

\hline
\end{tabular}
\end{center}
\label{tab:outflow_voff}
\end{table}

\begin{table}[ht]
\caption{[OIII] Outflow FWHM $[km/s]$ average in coronal line detections vs. non-detections}
\begin{center}
\begin{tabular}{c|c|c}
\hline
Line & CL Detections & CL Non-detections \\
\hline
	\lbrack \ion{Fe}{11}\rbrack$\lambda$7891 & $1208.71 \pm 0.00$ & $1118.68 \pm 490.62$ \\
	\lbrack \ion{Fe}{10}\rbrack$\lambda$6374 & $971.21 \pm 141.78$ & $1134.62 \pm 479.12$ \\
	\lbrack \ion{Fe}{7}\rbrack$\lambda$6087 & $984.82 \pm 393.22$ & $1153.02 \pm 481.21$ \\
	\lbrack \ion{Fe}{7}\rbrack$\lambda$5720 & $871.14 \pm 215.27$ & $1161.92 \pm 485.82$ \\
	\lbrack \ion{Ne}{5}\rbrack$\lambda$3426 & $1127.19 \pm 411.44$ & $1158.29 \pm 504.80$ \\
	\lbrack \ion{Ne}{5}\rbrack$\lambda$3345 & $1077.23 \pm 358.87$ & $1194.59 \pm 471.79$ \\
	\lbrack \ion{Ne}{4}\rbrack$\lambda$2424 & $1748.54 \pm 301.51$ & $1244.14 \pm 523.72$ \\

\hline
\end{tabular}
\end{center}
\label{tab:outflow_fwhm}
\end{table}

\section{Catalog Availability}
\label{sect:description}

The CLASS-Q catalog is available for download as a table in CSV or ASCII format. The catalog provides line properties for all detections and upper limits for all 19,508 Type~1 quasars analyzed in this work. Descriptions of each column in the table, as well as the units, data type, and source, are provided in Table \ref{tab:data_description} in the Appendix.  Note that the SDSS Spec Object ID column is a 64-bit integer, and must be read appropriately (either as a long integer, or a string) to avoid losing information. As in the CLASS catalog, all velocity offsets for coronal lines are measured relative to the stellar velocity, and all equivalent widths use a sign convention where emission is positive and absorption is negative.
\section{Discussion and Conclusions}

A key finding from this work is that optical CL emission is rare even in Type~1 quasars, given the sensitivity limit of SDSS. The rarity of coronal lines was also noted in the CLASS survey \citep{2023ApJS..265...21R}, however that sample was comprised of the general galaxy population, the vast majority of which do not have significant nuclear accretion activity. Here we focused exclusively on Type~1 quasars, in which a copious supply of high energy photons capable of producing the coronal line emission indisputably exists. In this work, we quantify for the first time the optical coronal line detection rate for a comprehensive set of optical lines in a large sample of Type~1 quasars, and find a detection fraction of only $\sim 4.5\%$ in SDSS. 

The most likely explanation of these findings is that dust suppresses line formation in the highly ionized gas surrounding AGNs. This explanation would be consistent with the recent finding that optical coronal line emission is preferentially detected in galaxies with the least dust extinction \citep{2023ApJ...945..127N}. The origin of the coronal line region has been debated for several decades, but a general consensus in recent years has emerged. Because the FWHM of the coronal lines are often in-between those of the strong lower ionization potential narrow lines and the broad permitted lines as noted in this and numerous previous works \citep{1991A&A...250...57A,2006ApJ...653.1098R,2011ApJ...743..100R,2017MNRAS.467..540L,2023ApJS..265...21R,2021ApJ...920...62N,2023ApJ...945..127N}, the CLs are thought to typically arise in gas that lies between the broad line region and the narrow line region traced by the lower ionization potential lines used in standard narrow line ratio diagnostics \citep[e.g.,][]{Baldwin1981}. This scenario might be expected given the higher critical densities associated with the CLs compared to those of the lower ionization potential lines. Indeed, the intermediate location of the coronal line region has been confirmed in the luminous Type~1 AGN NGC 3783 by recent VLT/GRAVITY observations of the [\ion{Ca}{8}] line \citep{GRAVITY2021}. These observations indicate that the line originates beyond both the broad line emitting clouds and the hot dust continuum imaged in the near-infrared, and therefore beyond the dust sublimation radius. However, extended coronal line emission with spatial extent from hundred pc to even kpc scales has also been observed \citep[e.g.,][]{2010MNRAS.405.1315M, 2011ApJ...739...69M, 2020ApJ...895L...9R, 2021ApJ...920...62N, 2023ApJ...945..127N, 2023MNRAS.524..143F}, possibly suggesting that radiative shocks can play a role in the production of the CL emission, although \citep{2024arXiv240815229M} argue that shocks are unlikely to be able to produced the observed CL luminosities, and that photoionization alone can produce extended CL emission.

In recent work, \citet{2024arXiv240815229M} explore the effects of dust on CL emission from gas ionized by an AGN. They find that the presence of dust has a dramatic effect on the coronal line emission, suppressing the line luminosity by as much as 3 orders of magnitude compared with dust-free gas. Dust has two key effects on line formation: (1) the emission lines are weaker because of gas-phase depletion onto dust grains, 2) dust absorbs ionizing radiation, suppressing line formation. Highly refractory elements such as Iron and Calcium are more severely affected by the presence of dust than the noble gases, such as Neon, providing a possible explanation for the rarity of the associated CLs from these species in the sample. Additionally, lower IP lines such as [\ion{O}{3}]~$\lambda$5007 are found to be less impacted by the presence of dust. Indeed, the models presented by \citet{2024arXiv240815229M}  indicate that the [Ne\,V]~$\lambda$3426 line is expected to be the most luminous line in dusty gas, consistent with the observations presented in this work.  The coronal line emitters in this sample of AGNs may be associated with AGNs in which grains are destroyed possibly as a result of winds generated by the central continuum source. Such a scenario would be consistent with the finding that the coronal line detection fraction appears to be higher in quasars that display outflows in the [\ion{O}{3}]~$\lambda$5007 line, and that the coronal lines themselves are often blueshifted and likely associated with energetic outflows \citep[e.g.,][]{1997A&A...323..707E,2006ApJ...653.1098R,2011ApJ...743..100R}. An association between CLs and outflows has previously been reported by \citet{2021ApJ...911...70B} in dwarf galaxies, suggesting that CL emission is enhanced in galaxies with outflows, possibly as a result of grain destruction in shocks.

Within the first year of observations, JWST has spectroscopically confirmed a growing number of galaxies with redshifts as high as $z\sim13.2$
\citep[e.g.,][]{2023A&A...677A..88B, 2023A&A...677A.115C, 2023NatAs...7..622C, 2023NatAs...7..611R, 2023ApJ...955...54S}. Some of these observations are revealing broad lines consistent with faint AGNs with black hole masses within the $10^6-10^7$\,$M_\odot$ range, all with narrow line ratios indistinguishable from low metallicity star forming galaxies \cite[e.g.,][]{2023arXiv230311946H, 2023ApJ...954L...4K, 2023ApJ...953L..29L, 2023arXiv230512492M, 2023arXiv230801230M, 2023A&A...677A.145U}. Many of the optical CLs presented in this work are redshifted into the wavelength range of JWST. Interestingly, none of the high-z broad line objects show prominent CL emission. Possible marginal detections of weak [\ion{Fe}{10}] $\lambda$6374 were reported by \citet{2023ApJ...954L...4K} in two $z>5$ sources in the CEERs program. Possible [\ion{Fe}{16}], [\ion{Ca}{5}], [\ion{Fe}{13}], and [\ion{Fe}{5}] are reported in a z = 5.55 galaxy by \citep{2023A&A...677A.145U}.  \citet{2023arXiv230512492M} report a marginal detection of the [\ion{Ne}{4}]$\lambda$2423 line in the $z=10.6$ galaxy, GN-z11. Interestingly, the luminosity of this reported line is approximately a factor of 7 lower than the average line luminosity reported in this low redshift sample of quasars. Finally, a prominent [\ion{Ne}{5}]$\lambda$3426 line is detected in GN42437, a low mass z=5.59 galaxy, possibly suggesting the presence of an intermediate mass black hole \citep{2024arXiv240218643C}. The line luminosity in this high redshift low mass galaxy is comparable to the lowest line luminosity found in this low redshift quasar catalog, and up to four orders of magnitude more luminous than the lowest line luminosities found in local dwarf galaxies in which the line is detected \citep{2021MNRAS.508.2556I,2023arXiv230502189H}. Interestingly, no other coronal lines are detected in the JWST spectrum of GN42437.

The faintness of the CLs in high-z targets may indicate that dust suppresses the line emission even in high-redshift sources, a result which may be consistent with recent findings of unexpectedly high dust obscuration in high-redshift sources based on SED modeling using deep JWST NIRCam and MIRI images\citep{2023MNRAS.518L..19R}, as well as the ubiquity of dust reddened AGNs revealed by JWST \citep{2023arXiv230905714G}. The CLASS-Q catalog presented here can provide a useful local benchmark of CL properties in Type~1 quasars that can be compared to the growing number of high-redshift spectra obtained by JWST.

\section{Acknowledgements}
JMC's work was supported by NASA through the CRESST II cooperative agreement under award number 80GSFC24M0006.

The simulations carried out in this work were run on ARGO and HOPPER, research computing clusters provided by the Office of Research Computing at George Mason University, VA. (\url{ http://orc.gmu.edu})

This research made use of Astropy,\footnote{\url{http://www.astropy.org}} a community-developed core Python package for Astronomy \citep{2013A&A...558A..33A}.  

 Funding for SDSS-III has been provided by the Alfred P. Sloan Foundation, the Participating Institutions, the National Science Foundation, and the U.S. Department of Energy Office of Science. The SDSS-III web site is \href{http://www.sdss3.org/}{http://www.sdss3.org/}.

 SDSS-III is managed by the Astrophysical Research Consortium for the Participating Institutions of the SDSS-III Collaboration including the University of Arizona, the Brazilian Participation Group, Brookhaven National Laboratory, Carnegie Mellon University, University of Florida, the French Participation Group, the German Participation Group, Harvard University, the Instituto de Astrofisica de Canarias, the Michigan State/Notre Dame/JINA Participation Group, Johns Hopkins University, Lawrence Berkeley National Laboratory, Max Planck Institute for Astrophysics, Max Planck Institute for Extraterrestrial Physics, New Mexico State University, New York University, Ohio State University, Pennsylvania State University, University of Portsmouth, Princeton University, the Spanish Participation Group, University of Tokyo, University of Utah, Vanderbilt University, University of Virginia, University of Washington, and Yale University. 

This publication makes use of data products from the Wide-field Infrared Survey Explorer, which is a joint project of the University of California, Los Angeles, and the Jet Propulsion Laboratory/California Institute of Technology, and NEOWISE, which is a project of the Jet Propulsion Laboratory/California Institute of Technology. WISE and NEOWISE are funded by the National Aeronautics and Space Administration.

\nocite{*}
\bibliographystyle{yahapj}
\bibliography{main}

\appendix

\startlongtable\begin{deluxetable}{ccccc}\tablewidth{0pt}\tablehead{\colhead{Column Name} & \colhead{Description} & \colhead{Units} & \colhead{Data Type} & \colhead{Source}}\startdata
SDSS\_NAME & SDSS Designation &  & string & 2\\
SpecObjID & SDSS Spec Object ID &  & 64-bit Integer & 2\\
RA & Right Ascension & degrees & float & 2\\
DEC & Declination & degrees & float & 2\\
Z & Cosmological redshift &  & float & 2\\
PLATE & SDSS Plate Number &  & integer & 2\\
MJD & SDSS Modified Julian Date of observation & MJD & integer & 2\\
FIBERID & SDSS Fiber ID &  & integer & 2\\
REST\_LAM\_MIN & Rest wavelength minimum & \AA & float & 2\\
REST\_LAM\_MAX & Rest wavelength maximum & \AA & float & 2\\
SHEN\_<column name> & Column from Shen+11 &  & float & 3\\
W1MPRO & WISE 1 profile-fit magnitude & mag & float & 4\\
W2MPRO & WISE 2 profile-fit magnitude & mag & float & 4\\
W3MPRO & WISE 3 profile-fit magnitude & mag & float & 4\\
W4MPRO & WISE 4 profile-fit magnitude & mag & float & 4\\
W1SIGMPRO & WISE 1 profile-fit magnitude error & mag & float & 4\\
W2SIGMPRO & WISE 2 profile-fit magnitude error & mag & float & 4\\
W3SIGMPRO & WISE 3 profile-fit magnitude error & mag & float & 4\\
W4SIGMPRO & WISE 4 profile-fit magnitude error & mag & float & 4\\
W12 & WISE 1 - WISE 2 color & mag & float & 4\\
W12\_ERR & Error in WISE 1 - WISE 2 & mag & float & 4\\
W34 & WISE 3 - WISE 4 color & mag & float & 4\\
W34\_ERR & Error in WISE 3 - WISE 4 & mag & float & 4\\
W1\_LOG\_FLUX & WISE 1 band flux & log(F/erg s$^{-1}$ cm$^{-2}$) & float & 4\\
W1\_LOG\_FLUX\_ERR & Error in WISE 1 band flux & log(F/erg s$^{-1}$ cm$^{-2}$) & float & 4\\
W2\_LOG\_FLUX & WISE 2 band flux & log(F/erg s$^{-1}$ cm$^{-2}$) & float & 4\\
W2\_LOG\_FLUX\_ERR & Error in WISE 2 band flux & log(F/erg s$^{-1}$ cm$^{-2}$) & float & 4\\
W3\_LOG\_FLUX & WISE 3 band flux & log(F/erg s$^{-1}$ cm$^{-2}$) & float & 4\\
W3\_LOG\_FLUX\_ERR & Error in WISE 3 band flux & log(F/erg s$^{-1}$ cm$^{-2}$) & float & 4\\
W4\_LOG\_FLUX & WISE 4 band flux & log(F/erg s$^{-1}$ cm$^{-2}$) & float & 4\\
W4\_LOG\_FLUX\_ERR & Error in WISE 4 band flux & log(F/erg s$^{-1}$ cm$^{-2}$) & float & 4\\
<Coronal Line>\_LOGF\_THRESH & The flux threshold of the spectrum at the coronal line & log(F/erg s$^{-1}$ cm$^{-2}$) & float & 1\\
<Coronal Line>\_LOGL\_THRESH & The luminosity threshold of the spectrum at the coronal line & log(L/erg s$^{-1}$) & float & 1\\
<Coronal Line>\_F\_FRAC & $\mathcal{F}$ metric of the coronal line &  & float & 1\\
<Coronal Line>\_NPIX & Number of continuous pixels above 3$\sigma$ for the coronal line &  & float & 1\\
<Coronal Line>\_SKY\_FLAG & \begin{tabular}[c]{@{}l@{}}Sky line proximity warning for the coronal line. True if within 20 $\AA$ of\\ 5578.5, 5894.6, 6301.7, or 7246.0 in the observed frame.\end{tabular} &  & boolean & 1\\
<Coronal Line>\_INSPECT & \begin{tabular}[c]{@{}l@{}}Visual inspection result. 1 if visually confirmed detection.\\-1 if visually confirmed nondetection. 0 if not visually confirmed.\end{tabular} &  & int & 1\\
<Coronal Line>\_LOG\_FLUX & Flux of the coronal line & log(F/erg s$^{-1}$ cm$^{-2}$) & float & 1\\
<Coronal Line>\_LOG\_FLUX\_ERR & Error in the flux of the coronal line & log(F/erg s$^{-1}$ cm$^{-2}$) & float & 1\\
<Coronal Line>\_LOGL & Luminosity of the coronal line & log(L/erg s$^{-1}$) & float & 1\\
<Coronal Line>\_FWHM & FWHM of the coronal line & km s$^{-1}$ & float & 1\\
<Coronal Line>\_FWHM\_ERR & Error in the FWHM of the coronal line & km s$^{-1}$ & float & 1\\
<Coronal Line>\_VOFF & Velocity offset of the coronal line & km s$^{-1}$ & float & 1\\
<Coronal Line>\_VOFF\_ERR & Error in the velocity offset of the coronal line & km s$^{-1}$ & float & 1\\
<Coronal Line>\_EW & Equivalent width of the coronal line & $\AA$ & float & 1\\
<Coronal Line>\_EW\_ERR & Error in the equivalent width of the coronal line & $\AA$ & float & 1\\
NA\_<Other Line>\_LOG\_FLUX & Flux of the Gaussian fit to the narrow component of the emission line & log(F/erg s$^{-1}$ cm$^{-2}$) & float & 1\\
NA\_<Other Line>\_LOG\_FLUX\_ERR & Error in flux of the narrow component of the emission line & log(F/erg s$^{-1}$ cm$^{-2}$) & float & 1\\
BR\_<Other Line>\_LOG\_FLUX & Flux of the Gaussian fit to the broad component of the emission line & log(F/erg s$^{-1}$ cm$^{-2}$) & float & 1\\
BR\_<Other Line>\_LOG\_FLUX\_ERR & Error in flux of the broad component of the emission line & log(F/erg s$^{-1}$ cm$^{-2}$) & float & 1\\
OUT\_<Other Line>\_LOG\_FLUX & Flux of the Gaussian fit to the outflow component of the emission line & log(F/erg s$^{-1}$ cm$^{-2}$) & float & 1\\
OUT\_OIII\_4960\_LOG\_FLUX\_ERR & Error in flux of the outflow component of the emission line & log(F/erg s$^{-1}$ cm$^{-2}$) & float & 1\\
<Other Line>\_LOG\_FLUX & Total flux of the emission line & log(F/erg s$^{-1}$ cm$^{-2}$) & float & 1\\
<Other Line>\_LOG\_FLUX\_ERR & Error in the total flux of the emission line & log(F/erg s$^{-1}$ cm$^{-2}$) & float & 1\\
<Other Line>\_LOGL & Luminosity of the emission line & log(L/erg s$^{-1}$) & float & 1\\
N\_CL\_DETECT & Number of coronal lines detected &  & int & 1\\
\label{tab:data_description}
\enddata
\begin{tablenotes}
    \item[0] Source references:
    \item[1] [1] This work
    \item[2] [2] SDSS
    \item[3] [3] Shen+11
    \item[4] [4] AllWISE
\end{tablenotes}
\end{deluxetable}
\(\) 

\end{document}